\documentclass[10pt,a4paper]{article}
\usepackage{amsfonts}
\usepackage{bm}
\usepackage{amssymb,amsmath}
\usepackage[algo2e]{algorithm2e}
\usepackage{algorithm}
\usepackage{algorithmic}
\usepackage{placeins}
\usepackage{color}
\usepackage{graphicx}
\newcommand{\RU}{\mathbb{R}}

\newcommand{\Ne}[1]{\mathbf{#1}}
\newcommand{\ca}[1]{\mathcal{#1}}

\newcommand{\tn}[1]{\textnormal{#1}}

\newcommand{\bvarphi}{{\boldsymbol{\varphi}}}
\usepackage{todonotes}
       
\newcommand{\comment}[1]{\vspace{5 mm}\par \noindent
  \marginpar{\textsc{Comment}} \framebox{\begin{minipage}[c]{0.95 
       \textwidth} \tt #1 \end{minipage}}\vspace{5 mm}\par}

\renewcommand{\comment}[1]{}

\begin{document}

\title{Variational integrators for the dynamics of thermo-elastic solids with finite speed thermal waves}

\author{
Pablo~Mata\\
\small Centro de Investigaci\'on en Ecosistemas de la Patagonia (CIEP),\\
\small Conicyt Regional/CIEP R10C1003, Universidad Austral de Chile.\\
\small Ignacio Serrrano 509, Coyhaique, Chile. 
\and 
Adri\'an~J.~Lew\\
\small Department of Mechanical Engineering, Stanford University,\\
\small Stanford,CA 94305-4040, USA\\
\small email:{lewa@stanford.edu}
}

\date{January, 2014}

%\cortext[Corauthor]{Durand 207, Department of Mechanical Engineering, Stanford University, Stanford,CA 94305-4040, USA.}

%
\maketitle
%\footnote{This work was partially founded by Conicyt Regional/CIEP/R10C1003.}\footnote{Department of the Army Research Grant, W911NF-07-2-0027.\\ National Science Foundation Career Award, CMMI-0747089.}

\begin{abstract}
This paper formulates variational integrators for finite element discretizations of deformable bodies with heat conduction in the form of finite speed thermal waves. The cornerstone of the construction consists in taking advantage of the fact that the Green-Naghdi theory of type II for thermo-elastic solids has a Hamiltonian structure. Thus, standard techniques to construct variational integrators can be applied to finite element discretizations of the problem. The resulting discrete-in-time trajectories are then  consistent with the laws of thermodynamics for these systems:  for an isolated system, they exactly conserve the total entropy, and nearly exactly conserve the total energy over exponentially long periods of time. Moreover, linear and angular momenta are also exactly conserved whenever the exact system does. For definiteness, we construct an explicit second-order accurate algorithm for affine tetrahedral elements in two and three-dimensions, and demonstrate its performance with numerical examples. 
\end{abstract}

%keywords: {Variational integrators, geometric integration, symplectic methods, thermo-elasticity.}
%
\tableofcontents
\section{Introduction}
The classical treatment to describe heat conduction in
solids adopts Fourier's law to describe the heat flux (e.g.,
\cite{Malvern1,Marsden1,Truesdell1}). In the context of thermo-elastic
materials, this assumption leads to the transmission of thermal disturbances with infinite
speed \cite{GurtinPipkin1,Chandrasekharaiah1}. Notwithstanding, this
model has been found to be very useful in many engineering
applications. On the other hand, since several decades ago
(e.g. \cite{DreyerAndStruchtrup}
%\footnote{A survey about the theory
% and experiments of heat pulses in dielectric materials at cryogenic
% temperatures.}
) it has been recognized that heat transport with
finite speed thermal waves may be useful in modeling some materials at low
temperatures. This phenomenon is frequently mentioned in the
literature as {\it second sound}, to differentiate it from the {\it
  first sound} in solids, which is related to the propagation of
mechanical waves.  Other contexts in which such theories could be
useful are referred to in 
\cite{CareyTsai1,ShenLittleHu1,ZhouaZhangaChen,Chester1}.

% Moreover, It has been also recognized that a hyperbolic theory of heat
% transport (i.e. one predicting finite speed thermal waves) may be
% useful even for describing thermal conduction at ambient temperature
% \cite{CareyTsai1}. In particular, recent technological developments in
% science and industries related to nanotechnology, laser-heating and
% cryogenic engineering require to consider non-Fourier type of heat
% conduction as noted in \cite{ShenLittleHu1,ZhouaZhangaChen} and
% reference therein. For example, Chester \cite{Chester1} computes a
% critical frequency for thermal fluctuations in dielectric solids
% assuming that heat transport proceeds by wave propagation rather than
% by diffusion, which is analogue to the second sound phenomenon in
% Helium II.

A number of theories have been proposed to model heat transfer with
finite speed thermal waves, e.g.,
\cite{GurtinPipkin1,LordShulman}. Among them, some incorporate a
regularization of the heat flux to include a relaxation
time controlling the velocity of propagation of heat perturbations
\cite{BogyNaghdi1,WallOlsson001}, see
\cite{IgnaczakAndOstojaStarzewski} for a thorough discussion.  A
variational theory for heat conduction in rigid solids with finite
wave speed based on a non-autonomous Hamiltonian was considered in
\cite{VujanovicBaclic1,VujanovicDjukic1,VujanovicStrauss001}. This
approach recovers the classical Fourier case in the limit in which a
relaxation parameter approaches zero. An attempt to extend these ideas to general
problems of non-conservative type is described in \cite{Vujanovic1},
and a survey of such models can be found in \cite{VujanovicJones001}.
Alternative variational principles are also available
\cite{Yang001,Gambar001,CannarozziUbertini,StanislawStephen}. More
recently, a general variational framework for non-reversible processes
was presented by Mielke and Ortiz \cite{MielkeOrtiz001} .

One of the most well-known formulations for heat-conducting
thermo-elasticity has been proposed by Green and Naghdi in a series of
papers
\cite{green1993thermoelasticity,GreenNaghdi1,GreenNaghdi2,GreenNaghdi3,GreenNaghdi100}. There, the authors take advantage of the concept of thermal
displacements (see, e.g., \cite{Maugin001a,Maugin001,Dascalu001} for
a discussion) to construct three types theories: (i) the first one (type I)
corresponds to the classical thermo-elasticity with heat conduction of
Fourier type, (ii) the second (type II) corresponds to a theory able to predict
the existence of undamped thermal waves traveling at finite speed, and
finally, (iii) the third (type III) results from combining the first
two types, see \cite{BargmannSteinmann105} for a recent discussion.

The Green and Naghdi theory of type II (G-N-II theory in what follows)
is particularly interesting due to the fact that: (i) It is well
suited for simulating second sounds in thermo-elastic
solids \cite{Chandrasekharaiah1,BargmannSteinmann101}, and (ii) the
resulting evolution equations possess an autonomous Hamiltonian
structure \cite{Maugin001}. This structure is unveiled after taking
into account that the thermal displacement, temperature and entropy
fields work in a complete analogy to how the (mechanical)
displacement, velocity and momentum fields do in an elastic
problem. In this case, the heat flux depends on the
gradient of the thermal displacements, and therefore, heat conduction
depends on the ``thermal strains.''  We will outline the
essential structure of this theory in the following sections.

A number of results are available for some versions of G-N-II. The
existence, uniqueness and the qualitative behavior of solutions for
the linear theory of thermoelasticity without energy dissipation in
the general anisotropic case has been studied by Quintanilla
\cite{Quintanilla001}.  Additionally, suitable conditions under which
the problem is well-posed are determined. In \cite{Quintanilla003} the
same author proposes a model extending the theory to the case of
non-simple elastic materials, by allowing the Helmholtz free energy
density of the material to depend on the second derivatives of the
displacements.  In \cite{QuintanillaStraughan001} a discussion about
G-N-III theory is presented. Chandrasekharaiah
\cite{Chandrasekharaiah100} studies uniqueness of solutions for
initial, mixed boundary value problems considering homogeneous and
isotropic materials.  In \cite{BargmannSteinmannJordan101} the
one-dimensional propagation of a Heaviside-shaped thermal pulse in a
half-space is invesigated analytically, and in
\cite{BargmannDenzerSteinmann101} the G-N theories are treated in the
context of the material force method.
  
The design and construction of numerical methods for thermo-mechanical
systems has a long history, see
e.g. \cite{armero1992new,FarhatParkDuboisPelerin,HolzapfelSimo1,SimoMiehe1,IbrahimbegovicChorfiGharzeddine}
to name only a few.  Outstanding among them are so called
fractional-step-methods, originally presented by Armero and Simo
\cite{armero1992new}, which are based on the splitting of the evolution
equations in an adiabatic phase followed by a conductive phase. The
resulting algorithms may be made unconditionally stable, but they are
only first-order accurate and, in general, implicit.
% Additionally,  the conservation of integrals of motion is not
% guaranteed. 
A goal in this area is to design {\it thermodynamically
  consistent} methods \cite{MataLew1,Romero1}, namely, methods whose
trajectories satisfy the first and second law of thermodynamics. This
means that the discrete trajectories should satisfy that, for an
isolated system, the energy of the system is conserved, and the
entropy of the system does not decrease in time. A step in this
direction is given by the extension of energy-momentum methods
\cite{Simo1,Simo2,Gonzalez001,Simo4} to the thermo-elastic case, see
e.g.,
\cite{GrossBetsch1,GrossBetsch2,KrugerGrobBetsch001,HeschBetsch001}
among others. A different approach to designing thermodynamically
consistent methods is followed by Romero
\cite{Romero1,RomeroI2}. Therein,  the author creates structure-preserving
discretizations of the so-called GENERIC formalism \cite{Ottinger001},
which conserve the invariants of the system even in the case of heat
conduction of Fourier type.

There is a shorter list of numerical methods for simulating second
sound. Bargmann and Steinmann
\cite{BargmannSteinmann101,BargmannSteinmann102,BargmannSteinmann103}
build a finite-element-based discretization in space and time. Moreover, they use the
resulting method to simulate the propagation of second sound at
cryogenic temperatures. An incremental formulation based in part on an
adequate discretization of the Euler-Lagrange (EL) equations is
presented in \cite{BargmannSteinmann104} by the same authors.

% In the last decades, atention of the researchers has been focused on developing methods able to capture more features of the dynamics. In particular, it is desirable to design methods predicting discrete flows which remain on the configuration manifold of the problem and simultaneously which are able to conserve the quantities asociated to symmetries of the system, among other characteristics (for a complete account about the construction of structure-preserving time integrators see e.g. \cite{Leimkuhler001,Hairer1,HairerEulerSymplectic}). 

The construction of integrators for which the trajectories preserve
invariants of the original system is the realm of structured or
geometric integration, see
e.g. \cite{Leimkuhler001,Hairer1,HairerEulerSymplectic}. Variational
integrators (VI) provide a way to design structure-preserving time
integrators for problems whose dynamics is generated by a
Hamiltonian. The basic idea behind these methods is to obtain the
algorithm from discrete analog to Hamilton's variational principle. In
this way, the computed discrete trajectories are stationary points of
a discrete action functional, the discrete action sum. As a result, the  discrete
trajectories approximate the exact trajectories of the system, and
nearly exactly preserve the exact energy of the system for long
times \cite[Ch. 9]{Hairer1}. Moreover, if the discrete action sum is designed
to respect symmetries of the original action functional, then by
virtue of a discrete version of Noether's theorem, the discrete
trajectories conserve the momenta conjugate to each one of these
symmetries, see e.g. \cite{Lew1}. Finally, the resulting algorithms
are symplectic by design. Nowadays it is possible to find a vast
literature on VIs. Most of the basic theory and
analytical results may be reviewed in
\cite{Veselov1,MoserV2,Kane001,Lew1,Lew2,Marsden2,Marsden3}.

These methods have been successfully applied in numerous fields, such
as to the construction of asynchronous integration methods in solid
mechanics \cite{Lew1,Lew2}, to problems with constraints
\cite{Marsden2,LeyendeckerMarsdenOrtiz101}, to problems with contact
\cite{FeMaOrWe,Cirak001,ryckman2010,harmon2009asynchronous}, with
oscillatory solutions \cite{stern2008implicit}, to Langevin and
stochastic differential equations
\cite{BouRabeeOwhadi101,bou2009stochastic}, to problems where the
evolution takes place over a nonlinear manifold \cite{Lee001,Lee002},
and to
incompressible fluids \cite{gawlik2011geometric}, to
name some of the most relevant ones.

In an earlier work \cite{MataLew1} we constructed VI for finite
dimensional thermo-elastic mechanical systems without heat
conduction. To construct such integrators we exploited the Hamiltonian
structure of such systems once the thermal displacements are
introduced. The resulting algorithms correspond to a family of
symplectic, entropy- and momentum-conserving, implicit and explicit
Runge-Kutta methods.  The formulation of
an explicit second-order method was particularly challenging in this
case, since the dependence of the Lagrangian on the thermal velocities
(temperatures) was other than quadratic. It turns out then that the
standard discrete Lagrangian  that leads to the central differences or
Newmark's second-order explicit algorithm renders an implicit
algorithm for these systems. Instead, we constructed a second-order
explicit algorithm by composing a first-order method with its adjoint.

In this paper, we extend those ideas to the case of a thermo-elastic
continuum according to G-N-II theory \cite{green1993thermoelasticity}.
We begin in \S \ref{ContinuumProblem} by reviewing the Lagrangian
formulation of the G-N-II theory, including a discussion about
symmetries and conserved quantities. The Euler-Lagrange equations of
this system are the balance of mechanical and thermal momenta. The
latter is none other than the evolution of the entropy density.  In
particular, the conservation of the total entropy
for isolated systems is a consequence of the invariance of the
Lagrangian density with respect to rigid translations of the thermal
displacements. This conservation property is added to the classical
conservation of energy and momenta appearing in (isolated) elastic
systems.

We construct a general class of time-integrators in \S
\ref{SectionDiscretization}. To this end, we first construct a
finite-dimensional mechanical system by introducing a finite element
discretization in space and a general class of quadrature rules.  In
particular, we only consider simplicial meshes of continuous $P_1$
finite elements. The construction of variational integrators then
follows in a standard way. The resulting integrators are generally
implicit.  We then construct explicit integrators in \S
\ref{ExplicitSecondOrder} by: (a) selecting two first-order time
integrators from the general class in \S \ref{SectionDiscretization},
the so-called {\it Euler-A} and {\it Euler-B} methods, and (b)
selecting a quadrature rule with quadrature points at the nodes of the
mesh (Gauss-Lobatto rule). Second-order accuracy follows by composing
these two methods, since they are mutually adjoint
\cite{Marsden2}. This leads to two different but related second-order
methods. A similar construction does not lead to explicit higher-order
methods, because of the order of the quadrature rule, so these are not
addressed here. The numerical performance of the algorithms is
examined in \S \ref{SectionNumericalExamples} through several
examples, and we finish the paper with some conclusions  in \S \ref{SectionConclusions}.

\section{Continuum problem}\label{ContinuumProblem}

In this section a Lagrangian formulation for the G-N-II of thermoelasticity \cite{green1993thermoelasticity} is presented
following the ideas described in
\cite{Dascalu001,Maugin001a,Maugin001,Maugin002}. The basic ingredient
of this construction is the inclusion of a {\it thermal
  displacement field} over the continuum, in addition to the
mechanical displacement field, and a free-energy that depends on the
``thermal strains.''  As a consequence, the balance equation
for the entropy of the system permits the propagation of thermal waves
of finite speed, due to the appearance of entropy fluxes (or ``thermal
stresses'') that depend on
the gradient of the thermal displacements.
%
%--------------------------------------------------------------------------------------------------------------
%
\subsection{Kinematics and constitutive relations}

Consider a body with reference configuration ${\Omega}\subset\RU^d$,
$d\in\{1,2,3\}$, where $\Omega$ is a bounded open set with piecewise
smooth boundary $\partial\Omega$\footnote{For simplicity,
  polyhedral, so that it can be exactly meshed, although curved
  domains can be handled in a ``natural'' way.}. The deformation of the body is described by a motion
$\bvarphi(\Ne{X},t)$, which returns the position in $\RU^d$ at time
$t$ of the continuum particle at $\Ne X\in \Omega$. Similarly, consider the thermal displacement field
$\Phi(\Ne X,t)$, a real-valued function which in the context of this
theory is an additional kinematic variable
\cite{Dascalu001,Maugin001,Maugin002}. A thermo-elastic deformation of
the body is described then as $(\bvarphi, \Phi)$. The spatial gradient
of a thermo-elastic deformation are the deformation gradient $\Ne
F=\nabla \bvarphi$ and the thermal displacement gradient $\bm{\beta}=\nabla \Phi$. 
The temporal derivatives include the velocity
field $\bm v=\dot \bvarphi$ and the {\it empirical temperature} field
$\theta = \dot \Phi$, which is the distinguishing feature of the
thermal displacements.

Following Green-Naghdi's theory of type II for non-classical
thermo-elasticity \cite{Dascalu001, GreenNaghdi1, green1993thermoelasticity}, we
assume that the Helmholtz free energy density per unit of mass is a
smooth real-valued, material-frame indifferent function
\begin{equation}
  \tn{\textsf{A}}(\Ne{F},\bm{\beta},\theta;\Ne{X}).  
\end{equation}
Its distinctive feature is that it depends on $\bm\beta$, the gradient of
the thermal displacements. The function $\tn{\textsf{A}}$ serves then
as the thermodynamic potential for the first Piola-Kirchhoff stress
tensor $\Ne{P}$, the entropy density per unit mass $\eta$, and the
entropy flux vector (per unit area in the reference configuration) $\Ne{h}$, as follows
\begin{equation}\label{ConstitutiveEquations}
  \Ne{P}=
  \rho_0\dfrac{\partial\tn{\textsf A} }{\partial\Ne F},
  \qquad\eta=
  -\dfrac{\partial\tn{\textsf A}}{\partial\theta},
  \qquad\Ne{h}=
  -\rho_0\dfrac{\partial\tn{\textsf A}}{\partial\bm\beta},
\end{equation}
where $\rho_0(\Ne X)$ is the mass density per unit volume in  the
reference configuration. If, for example, $\textsf A$ is strictly convex with respect to $\theta$,
it is possible to invert (\ref{ConstitutiveEquations})$_2$  for each
value of $\Ne F$ and $\bm\beta$ to compute the temperature as
\begin{equation}\label{Temperature}
  \theta=\hat{\theta}(\Ne{F},\bm \beta,\eta;\Ne X).
\end{equation}
The heat flux vector $\Ne{q}$ (per unit area in the
reference configuration) follows from the constitutive relation (\ref{ConstitutiveEquations})$_3$ as
\begin{equation}
  \Ne{q}=\theta\Ne{h}.
\end{equation}
Finally, for future reference we also introduce the internal energy
density per unit mass as
\begin{equation}
  \label{eq:8}
  \textsf U(\Ne F, \bm \beta, \eta; \Ne X) = \eta\; \hat{\theta}(\Ne{F},\bm
  \beta,\eta;\Ne X) +\textsf A(\Ne F, \bm \beta,  \hat{\theta}(\Ne{F},\bm
  \beta,\eta;\Ne X); \Ne X).
\end{equation}
In the following we shall assume that the material is spatially
homogeneous, namely, that $\textsf A$ and hence $\textsf U$ do not
depend explicitly on $\Ne X$. We also assume that $\textsf A$ is a
strictly convex function of $\theta$ for all possible values $(\Ne
F,\bm\beta)$. Additionally, to obtain an explicit algorithm, we assume
that $\partial^2 \textsf A/\partial \bm\beta\partial \theta=0$, i.e.,
there is no dependence of $\eta$ on $\bm \beta$ and of $ {\Ne h}$ on
$\theta$. This
assumption is satisfied by standard models in the literature for this
type of heat conduction.

\comment{AL:
 In particular, this implies that the function $\hat \theta$
in \eqref{Temperature} does not depend on $\bm \beta$ either.
This follows from 
\begin{displaymath}
  \eta = -\frac{\partial A}{\partial \theta} (\Ne F, \bm \beta, \hat
  \theta(\Ne F, \bm \beta, \eta))
\end{displaymath}
for all $(\Ne F,\bm\beta,\eta)$, so
\begin{displaymath}
  0 = - \underbrace{\frac{\partial^2A}{\partial \theta\partial\bm \beta}}_{=0} -
  \underbrace{\frac{\partial^2 A}{\partial \theta^2}}_{>0,\text{ by convexity}} \frac{\partial \hat \theta}{\partial
    \bm \beta}.
\end{displaymath}

}

%
%------------------------------------------------------------------------------------------------------
%
\subsection{Lagrangian formulation}\label{LagrangianFormulation}

A unique feature of this theory is that the local form of the
balance equations for the stresses and the entropy density are the
Euler-Lagrange equations of  Hamilton's principle with a suitable
action functional. To this end, we define the appropriate Lagrangian as
\begin{equation}\label{LagrangianFunctional}
  \tn{\textsf{L}}(\bm\varphi,\Phi,\bm v,\theta):=
  \tn{\textsf{K}}(\bm v)-\tn{\textsf W}(\bm{\varphi},\Phi,\theta)-
  \tn{\textsf{E}}(\bm{\varphi},\Phi),
\end{equation} 
where $\tn{\textsf{K}}(\bm v)$ corresponds to the kinetic energy and it is given by
\begin{equation}\label{KineticEnergy}
  \tn{\textsf K}(\bm v)=
  \dfrac{1}{2}\int_{\Omega}\rho_0\bm v\cdot\bm v\; dV,
  \qquad
\end{equation}
the function $\tn{\textsf W}(\bm{\varphi},\Phi,\theta)$ is the total Helmholtz free energy of the body,
\begin{equation}\label{InternalPotential1}
  \tn{\textsf W}(\bm{\varphi},\Phi,\theta)=
  \int_{\Omega}\rho_0\tn{\textsf{A}}(\nabla \bvarphi,\nabla \Phi,\theta)\; dV,
\end{equation}  
and the remaining term $\tn{\textsf{E}}(\bm{\varphi},\Phi)$ stems from the action of external agents acting on the
system. For this term we assume the existence of a conservative body
force per unit mass $\Ne B=-d V_B(\bvarphi)/d\bvarphi$, for a
potential function $V_B$. Additionally,
we consider an entropy source per unit mass $\tn{\textsf{Q}}\colon
\Omega\rightarrow \RU$, and a traction field $\Ne T:\Gamma_{\tn
  t}\subset\partial\Omega\rightarrow\RU^d$ and an entropy flux into
the body
$\overline{\tn h}:\Gamma_{\tn h}\subset\partial\Omega\rightarrow\RU$ (both per unit area in the reference configuration). Under these
conditions, we have
\begin{equation}
  \label{ExternalPotential2}  
  \tn{\textsf{E}}(\bm{\varphi},\Phi)=
  \int_{\Omega}\rho_0 \left(V_B(\bvarphi)- \tn{\textsf Q}\; \Phi\right)\;
  dV 
  -\int_{\Gamma_{\tn t}}\Ne T\cdot \bvarphi\; dS 
  -\int_{\Gamma_{\tn{h}}}\overline{\tn h}\; \Phi\; dS.
\end{equation}

To formulate Hamilton's principle, consider the set $\ca{C}$ of
smooth enough motions $(\bm\varphi,\Phi)$ during the time interval
$[0,T]$ between two prescribed configurations
$(\bm\varphi,\Phi)\big\vert_{t=0}=(\bm{\varphi}_0,\Phi_0)$ and
$(\bm\varphi,\Phi)\big\vert_{t=T}=(\bm{\varphi}_T,\Phi_T)$. Motions in
$\ca C$ are assumed to satisfy boundary
conditions  on $\Gamma_\varphi=\partial \Omega \setminus \Gamma_t$ and
$\Gamma_\Phi=\partial \Omega\setminus \Gamma_h$ according to
\begin{equation}\label{Dirichlet_1}
 \bm{\varphi}(\cdot, t)\big\vert_{\Gamma_\varphi}=\overline{ \bm\varphi }(\cdot,t)
 \quad\tn{and}\quad
 \Phi(\cdot,t)\big\vert_{\Gamma_\Phi}=\overline{ \Phi }(\cdot,t)
\end{equation} 
for all $t\in(0,T)$, where $\overline \bvarphi\colon \Gamma_\varphi\times(0,T)\rightarrow
\RU^d$ and $\overline \Phi\colon \Gamma_\Phi\times(0,T)\rightarrow
\RU$ are prescribed.
Then, the action functional $\tn{\textsf{S}}:\ca{C}\rightarrow\RU$ is given by
\begin{equation}\label{ActionFunctional}
  \tn{\textsf S}\left(\bm{\varphi},\Phi\right)=
  \int_0^T\tn{\textsf{L}}(\bm{\varphi},\Phi,\dot{\bm{\varphi}},\dot{\Phi})\; dt.
\end{equation}

Hamilton's principle states that the trajectory followed by the system
corresponds to a stationary point of the action in $\ca C$. In other
words, the problem of studying the motion of a thermo-elastic system
consist in finding  $(\bm{\varphi},\Phi)\in \ca{C}$ such that 
$$\delta\tn{\textsf{S}}=
%\frac{\tn{d}}{\tn{d}\epsilon}\tn{\textsf{S}}(\bm{\varphi}+\epsilon\delta\bm{\varphi},\Phi+\epsilon\delta\Phi)\big\vert_{\epsilon=0^+}=
0,$$ 
for all admissible variations
$(\delta\bm{\varphi},\delta\Phi)$ (in $\textsf T\ca{C}$).

Therefore, taking the first variation of (\ref{ActionFunctional}),
integrating by parts in time and applying the divergence theorem in
the space variables yields 
\begin{eqnarray}
  \delta\tn{\textsf S}
  &=&
  \int_0^T\int_{\Omega}\Big[(\nabla\cdot\Ne{P}+\rho_0 \Ne B-\rho_0\ddot{\bm{\varphi}})
  \cdot\delta\bm{\varphi}+
  (\rho_0
  \tn{\textsf{Q}}-\nabla\cdot\Ne{h}-\rho_0\dot{\eta})~\delta\Phi\Big]\;
dV\; dt +
  \nonumber\\
  &&
  +\int_0^T\left[\int_{\Gamma_{\tn t}}(\Ne
    T-\Ne{P}\Ne{n})\cdot\delta\bm{\varphi}\; dS+
   \int_{\Gamma_{\tn{h}}}(\overline{\tn
     h}+\Ne{h}\cdot\Ne{n})~\delta\Phi\; dS\right]\;dt=0,
\end{eqnarray}
where $\Ne n$ is the outward normal to $\partial \Omega$. Then, the following system of Euler-Lagrange equations are obtained
\begin{subequations}
\begin{alignat}{3}
  \rho_0\ddot{\bm{\varphi}}
  &=\nabla\cdot\Ne{P}+\rho_0\Ne B, &\qquad\tn{in} &\qquad \Omega\times[0,T], 
  \label{Balance1}
  \\
  \rho_0\dot{\eta} 
  &=\rho_0
  \tn{\textsf{Q}}-\nabla\cdot\Ne{h}, &\qquad\tn{in} &\qquad\Omega\times[0,T],
  \label{Balance2}
  \\
  \Ne{P}\Ne{n}
  &=
  \Ne T, &\qquad\tn{on} &\qquad\Gamma_{\tn t}\times[0,T],
  \label{Balance3}
  \\
  -\Ne{h}\cdot\Ne{n}
  &=
  \overline{\tn h}, &\qquad\tn{on} &\qquad\Gamma_{\tn{h}}\times[0,T].
  \label{Balance4}
\end{alignat}
An initial value problem is obtained by complementing these equations
with  the boundary conditions (\ref{Dirichlet_1}) and 
%
%\begin{alignat}{3}\bm{\varphi}&=
%\overline{\bm\varphi}, &\qquad\tn{on} &\qquad\Gamma_\varphi\times[0,T],\label{Balance5_a}
%  \\ \Phi  &= \overline{\Phi}, &\qquad\tn{on} &\qquad\Gamma_{\Phi}\times[0,T],\label{Balance5_b}
%\end{alignat}
%
initial conditions
\begin{equation}\label{Balance6}
  (\bm{\varphi},\Phi)\big\vert_{t=0}=
  (\bm{\varphi}_0,\Phi_0)
  \qquad\tn{and} \qquad
  (\dot{\bm{\varphi}},\theta)\big\vert_{t=0}=
  (\pmb{v}_0,\theta_0)
  \qquad\tn{in} \qquad\Omega.
\end{equation}
\end{subequations}
Some remarks are now pertinent:
\begin{enumerate}
\item[(i)] The local balance of linear momentum is stated in
  (\ref{Balance1}). Additionally, the entropy balance equation
  (\ref{Balance2}) appears as a consequence of the variational
  principle. By expanding its left hand side the following
  mechanically-coupled heat equation is obtained
\begin{equation*}
  \dfrac{\partial\eta}{\partial\Ne F}\bm{:}\dot{\Ne F}+
  \dfrac{\partial\eta}{\partial\theta}\dot{\theta}=
  \tn{\textsf Q}-
  \frac{1}{\rho_0}\nabla\cdot\Ne h,
\end{equation*}
in which time derivatives of  only $(\bvarphi,\Phi)$ appear. 
\item[(ii)] These equations are formally identical to  those for
  thermo-elasticity based
  on Fourier's law. The key difference lies in that here $\Ne h$
  depends on $\nabla \Phi$, while in the classical case $\Ne h$
  depends on $\nabla \dot \Phi$.

\item[(iii)] From Clausius-Duhem inequality  the internal rate of
  entropy production should satisfy
  \begin{equation}
    \label{eq:1}
    \zeta = \rho_0 (\dot \eta - \tn{\textsf Q}) + \nabla \cdot\Ne
    h\ge 0.
  \end{equation}
  From \eqref{Balance2}, it follows that $\zeta=0$ even in the
  presence of heat conduction. This why this theory is referred to as
  {\it thermo-elasticity without energy dissipation} \cite{green1993thermoelasticity,GurtinPipkin1}.

%\item[(iii)] As commonly done, if viscous dissipation or In many practical cases the externally applied mechanical and thermal forces can not be derived from a potential due to the fact that they are intrinsically non-conservative. If this is the case, Hamilton's principle is still valid even when it is not longer a variational statement but rather a weak form analogous to D'Alembert principle in classic mechanics \cite{Marsden4}. %The way in which this principle is built consists in replacing the term corresponding to the external component (\ref{ExternalPotential}) by that due to the work done by non-conservative mechanical and thermal forces acting on the system. 
%

\end{enumerate}

%----------------------------------------------------------------------------------------------------
\subsection{Conserved quantities}\label{ConservedQuantitiesContinuum}
According to Noether's theorem (see, e.g., \cite{Marsden1,Marsden3})
if the Lagrangian of the system remains invariant under the action of
a group on the configuration space, there exists a corresponding
quantity which is conserved by the dynamics. In this section,
conserved quantities are reviewed for the G-N-II theory of
thermo-elasticity.  For simplicity, in this section we assume that 
  $\Gamma_\bvarphi=\Gamma_\Phi=\emptyset$. 

\paragraph{Energy conservation} The Lagrangian functional
(\ref{LagrangianFunctional}) is autonomous and therefore invariant
under translations in time (some times called horizontal translations,
see e.g. \cite{ThesisAdrian,Marsden2} and references therein). It is
well known that for such systems the associated conserved quantity is
the value of the Hamiltonian function, which is that of the total energy. 
%(see(\ref{HamiltonianFuntional1}) in Appendix \ref{App1})
 An alternative
way to see this is by direct computation of the time derivative of the
Hamiltonian function, verifying that it is exactly equal to zero if
the Euler-Lagrange equations (\ref{Balance1})--(\ref{Balance4}) are
satisfied.

\paragraph{Linear and angular momentum conservation} If
$\Ne B=0$, $\tn{\textsf Q}=0$,  $\Ne T=0$ and $\overline{\tn h}=0$,  the Lagrangian (\ref{LagrangianFunctional}) is
invariant under the action of rigid body translations and rotations in
space. A well-known consequence is that the linear and angular
momentum of the system are conserved in time.

\paragraph{Conservation of the total entropy} Conservation of energy
and linear and angular momentum are features shared with classical
elastodynamics. The hallmark feature of this theory is that, in the
absence of entropy sources ($\tn{\textsf Q}=0$ and $\overline{\tn
  h}=0$), the entropy of the system does not increase even in the
presence of heat conduction.

A simple way to see this is to integrate (\ref{Balance2}) over
$\Omega$ and apply the divergence theorem to obtain the time
derivative of the total entropy $\Xi$, namely,
\begin{equation}
  \dot\Xi:=\int_\Omega\rho_0\dot{\eta} \; dV= -
  \int_{\partial\Omega}\Ne{h}\cdot\Ne n\; dS=0.
\end{equation}

An alternative perspective on this conservation law is as a symmetry
of the Lagrangian. More precisely, the total entropy appears as the
conserved quantity associated to the invariance of the Lagrangian
under the continuous transformation of the thermal displacements
$\Phi_{\epsilon}\equiv\Phi+\epsilon \zeta$ for and $\zeta\in \RU$ and
any $\epsilon\in\RU$. This symmetry is stated as
\begin{equation}
  \tn{\textsf{L}}_{\epsilon}\equiv
  \tn{\textsf{L}}(\bm{\varphi},\dot{\bm{\varphi}},\nabla
  \Phi_{\epsilon},\dot \Phi_\epsilon)=
  \tn{\textsf{L}}(\bm{\varphi},\dot{\bm{\varphi}},\nabla\Phi_0,\dot \Phi_0)=
  \tn{\textsf{L}}_0,
\end{equation}
after noting that $\nabla\Phi_{\epsilon}=\nabla\Phi$ and
$\dot{\Phi}_{\epsilon}=\dot{\Phi}$. 
The conserved quantity associated to this symmetry can be unveiled by computing
\begin{equation}
0=  \dfrac{\tn{d}}{\tn{d}\epsilon}\tn{\textsf{L}}_{\epsilon}\Big\vert_{\epsilon=0}
  =
  \int_\Omega\left( \rho_0 \eta \frac{\tn d}{\tn d t} 
\zeta +
  \nabla\cdot\Ne{h}\; \zeta
\right) \; dV
  =
  \int_\Omega\left( \rho_0 \eta \frac{\tn d}{\tn d t} 
\zeta -
  \rho_0 \dot \eta \; \zeta
\right) \; dV
= \zeta\; \frac{\tn d}{\tn dt}
  \int_\Omega\rho_0 \eta \;  dV,
\end{equation}
where we have used (\ref{Balance2}) and (\ref{Balance4}). The fact
that $\dot\Xi=0$ follows by setting
$\zeta\not=0$.

If the body is in thermal contact with the surrounding environment, we have that the entropy source $\tn{\textsf{Q}}$ and entropy flux $\overline{\tn h}$ modify the above equation according to
\begin{equation}\label{EntropyRate001}
  \dot{\Xi}=\int_{\Omega}\tn{\textsf{Q}}\; dV+
  \int_{\Gamma_{\tn{h}}}\overline{\tn{h}}\; dS,
\end{equation}
which clearly shows that the time evolution of the total entropy in the system depends on the signs and magnitudes of the external sources and fluxes. 

 %In particular, for positive net fluxes the entropy of the system is strictly non decreasing.    
%
%\item[(ii)] The classical theory of heat conduction with Fourier law and to the relation

%\begin{equation}\rho\theta\xi =-\Ne h\cdot\nabla\theta,\qquad\Ne q=-k\nabla\theta,\end{equation}
%then
%\begin{equation}  \rho\dot{\eta} =   \rho s-\frac{\nabla\cdot\Ne q}{\theta}=  \rho s-\nabla\cdot\Ne h + k^{-1}\Ne h\cdot\Ne h.\end{equation}
% It is frequent to replace in the previous equation $\theta$ by $\theta_{ref}$ to obtain a linear equation as\begin{equation}\rho\dot{\eta} = \rho s-\kappa^\ast\nabla\cdot\nabla\theta\end{equation}
%where $\kappa^\ast=\kappa\theta_{ref}^{-1}$.
%
%\todo{AL: Include something about the internal rate of entropy production}

%----------------------------------------------------------------------------------------------------------
%
\subsection{Examples}\label{Example1Continuum}
In this section we show two examples of constitutive relations for
G-N-II thermo-elastic  materials.

\paragraph{Example of linear thermo-elasticity} A possible form for
the Helmholtz free energy density per unit mass for the case of infinitesimal deformations is given by
\begin{equation*}
  \tn{\textsf A}(\Ne F,\bm\beta,\theta)=
  \dfrac{1}{2\rho_0}\bm e \bm :\Ne C \bm : \bm e-
  \dfrac{c}{2\theta_0}(\theta-\theta_0)^2-
  \gamma(\theta-\theta_0)\bm e\bm :\Ne I-
  (\theta-\theta_0)\eta_0+
  \dfrac{\kappa}{2\rho_0}\bm{\beta}\cdot\bm{\beta},
\end{equation*}
where the operator $\bm{:}$ indicates double contraction of indices,
$\bm{e}=\frac 1 2(\Ne F+\Ne F^{\tn t}-2\Ne I)$ is the small strain
tensor, $\Ne{C}$ are the linear elastic moduli (assumed to have major
and minor symmetries), $c$ is a specific heat, $\gamma$ is a
thermo-mechanical coupling parameter, $\kappa$ is a non-classical
thermal conductivity constant, and $\theta_0$ and $\eta_0$ are the
reference temperature and reference entropy, respectively.

Therefore, the  entropy flux vector, the internal entropy density, the temperature, and the Cauchy stress tensor  (which coincides with
$\Ne{P}$ in the infinitesimal deformations case) are given by
\begin{subequations}
\begin{eqnarray}  
  \Ne{h}
  &=&
  -\kappa\bm{\beta},
  \\
  \eta
  &=&
  \dfrac{c}{\theta_0}(\theta-\theta_0)+
  \gamma\bm e\bm :\Ne I+\eta_0, 
  \label{entropy1}
  \\
  \theta
  &=&
  \theta_0
  \left(
  1+  
  \dfrac{\eta-\eta_0-\gamma\bm e\bm :\Ne I }{c}
  \right),  
  \\
  \bm \sigma
  &=&
  \Ne C\bm :\bm e-\rho_0\gamma(\theta-\theta_0)\Ne I.
\end{eqnarray}
\end{subequations}
The corresponding balance equations are given by
\begin{eqnarray*}
\nabla \cdot(  \Ne{C}\bm{:}\bm e)-
\rho_0\gamma\nabla\theta
  &=&  
  \rho_0(\ddot{\bm\varphi}-\Ne B),
  \\
  \dfrac{c}{\theta_0}\dot{\theta}+
  \gamma\dot{\bm e}\bm :\Ne I
  &=&
  \tn{\textsf Q}+\frac{\kappa}{\rho_0}\nabla\cdot\bm\beta.
\end{eqnarray*}
%
%---------------------------------------------------------------------------------------------
%

\paragraph{One-dimensional harmonic solutions:} It is interesting to
find one-dimensional harmonic solutions of these equations, since they
reveal coupled thermo-mechanical waveforms that propagate without
distortion. 
Assuming that the material is isotropic, that $\Phi(\Ne X,t)\equiv \Phi(X,t)$,
$\bm\varphi(\Ne X,t)=\Ne X+ u(X,t) \Ne e_X$, where $\Ne e_X$ is the
unit vector in the $X$-direction, and in the absence of external
loading ($\Ne B=0$, $\textsf Q=0$), the above equations reduce to
\begin{subequations}
\label{eq:30}  
\begin{eqnarray}
  \ddot{u} 
  &=&
  \Big(\dfrac{\tn{E}}{\rho_0}\Big)u,_{XX}-
  \gamma\dot{\Phi},_{X}, \label{eq:31}
  \\
  \ddot{\Phi}
  &=&
  \Big(\dfrac{\kappa\theta_0}{c\rho_0}\Big)\Phi,_{XX}-
  \Big(\dfrac{\gamma\theta_0}{c}\Big)\dot{u},_X. \label{eq:32}
\end{eqnarray}
\end{subequations}
Here  $E$ is the effective stiffness of the material in this case.  We
consider traveling wave solutions of the type $u(X,t)=\Re[{A_\varphi \exp[i(K
X+\omega t)]}]$ and $\Phi(X,t)=\Re[{A_\Phi\exp[i(K X+\omega t)]}]$,
which after replacing in the above equations yield the following four dispersion relations
\begin{equation}\label{OmegaLinear}
\omega_{\pm\pm}=
\pm K 
\sqrt{
\frac{
\theta_0(\rho_0\gamma^2+\kappa)+c E
\pm
\sqrt{
\rho_0^2\theta_0^2\gamma^4+
2 \rho_0 \theta_0 \gamma^2 \left( 
c E+\kappa\theta_0
\right)+
\left(
cE-\kappa\theta_0
\right)^2
}
}{2c\rho_0}
}.
\end{equation}
%
%In particular, the the constant $E\rho_0^{-1}$ correspond to the velocity of mechanical waves, which is frequently quoted as {\it first sound} and $\kappa\theta_0 c^{-1}$ is the velocity of propagation of thermally induced waves or {\it second sound} . 
%
%
A special case of this dispersion relation is obtained by setting
$\gamma=0$ (no thermo-mechanical coupling), from where we recover the dispersion relations for the thermal and mechanical waves \cite{BargmannSteinmann101,BargmannSteinmann102}  
$$\dfrac{d\omega}{dK}=
\pm\sqrt{
\frac{E}{\rho_0}
}
\qquad\tn{and}\qquad
\dfrac{d\omega}{dK}=
\pm\sqrt{
\frac{\kappa\theta_0}{c\rho_0}
},
$$
respectively. Moreover, if we consider the case $\kappa=0$ and
$\gamma>0$, we obtain the dispersion relation corresponding to an
adiabatic thermo-mechanical wave, namely,
$$\dfrac{d\omega}{dK}=\pm
\sqrt{
\frac{\rho_0\theta_0\gamma^2+cE}{c\rho_0}
}.$$
%    

%------------------------------------------------------------------------------------------------
%
\paragraph{Example in nonlinear thermo-elasticity} One form for the
Helmholtz free energy density per unit mass in the large deformations case is given by
\begin{eqnarray}\label{NonLinearConstitutiveEquation}
  \tn{\textsf{A}}(\Ne F,\bm\beta,\theta)
  &=&
  \dfrac{\mu}{2\rho_0} \Ne{F}\bm : \Ne{F}+
  \dfrac{\lambda}{2\rho_0}\ln^2 J-
  \dfrac{\mu}{\rho_0}\ln J-
  \gamma(\theta-\theta_0)\ln J
  \nonumber
  \\
  &&
  +c\left(\theta-\theta_0-\theta\ln\Big(\dfrac{\theta}{\theta_0}\Big)\right)-
  (\theta-\theta_0)\eta_0+
  \frac{\kappa}{2\rho_0}\bm{\beta}\cdot\bm{\beta},
\end{eqnarray}
where $\gamma$, $\kappa$, $\rho_0$ and $c$ are defined as in the
previous example, $J=\tn{det}(\Ne{F})$, and $\lambda$, $\mu>0$ are
elastic  constants. It follows from here that
\begin{subequations}
\begin{eqnarray}  
  \Ne{h}
  &=&
  -\kappa\bm{\beta},
  \\
  \eta
  &=&
  c\ln\Big(\dfrac{\theta}{\theta_0}\Big)+
  \gamma\ln J+\eta_0, 
  \label{entropy2}
  \\
  \theta
  &=&
  \theta_0\tn{exp}
  \left[
  \frac{\eta-\eta_0-\gamma\ln J}{c}
  \right],  
  \\
  \Ne{P}
  &=&
  \mu\Ne{F}+\left(\lambda\ln J-
  \mu-\rho_0\gamma(\theta-\theta_0)\right)\Ne{F}^{-\sf T}.
\end{eqnarray}
\end{subequations}
%
% with balance equations
% %
% \begin{subequations}
% \begin{eqnarray}
%   \rho_0\ddot{\bm{\varphi}}
%   &=&
%   \Big(
%   \lambda\Ne{F}^{-\sf T}\bm{:}
%   \nabla\Ne{F}-\rho_0\gamma\nabla\theta\Big)\cdot\Ne{F}^{-\sf{T}}
% %  \nonumber
% %  \\
% %  &&
%   +\Big(
%   \lambda\ln J-\mu-\rho_0\gamma(\theta-\theta_0)
%   \Big)
%   \Ne{D}\bm{:}\nabla\Ne{F}+\mu\nabla\cdot\Ne{F}+\rho_0\Ne B, 
%   \label{Balance1Eje1}
%   \\
%   c\theta_0\frac{\dot\theta}{\theta}
%   &=&
%   \tn{\textsf Q}-
%   \gamma\Ne{F}^{-\tn{t}}\bm{:}\dot{\Ne{F}}+
%   \frac{\kappa}{\rho_0}\nabla\cdot\bm\beta,
%   \label{Balance1Eje2}
% \end{eqnarray}
% \end{subequations}
% %
% where $\tn{D}_{ijkl}:=\tn{F}^{-\tn{t}}_{li}\tn{F}^{-\tn{t}}_{jk}$.
%
%-------------------------------------------------------------------------------------------------------
%
\section{Discretization}\label{SectionDiscretization}

We next construct new variational integrators to approximate the motions
of bodies made of thermo-elastic material obeying the G-N-II theory of
heat conduction. We accomplish this by following the same steps as in
\cite{Lew2} for nonlinear elasticity. We first introduce a finite element
discretization of the body in space
 and obtain semi-discrete equations
of evolution in time for its degrees of freedom. Variational
integrators in time are then constructed for the resulting finite
dimensional system. 

The variational integrators here constructed enjoy a number of
important properties which will be described in detail in the following
sections: (i) they are symplectic, (ii) they exactly
conserve linear and angular momentum whenever they have to be
conserved, (iii) they ensure the exact conservation of the total
entropy for thermally isolated systems, and (iv) they display a bounded
grow in the error of the computed energy which remains
close to the exact value for exponentially long periods of
time. 
% Moreover if the body is in thermal contact with the environment
% through entropy sources, the algorithms guarantee that the
% entropy changes . \todo{AL: Have we defined what
%   thermodynamically consistent means?}   

%
%--------------------------------------------------------------------------------------------------------------
%

\subsection{Discretization in space}\label{sec:discretization-space}

Consider a triangulation $\ca T_h$ of ${\Omega}$ with $N$ nodes, where $h$ is the
maximum diameter of an element in $\ca T_h$, in which $\Gamma_t$ and
$\Gamma_h$ are exactly meshed by the restriction of $\ca T_h$ to
$\partial \Omega$. We focus here on first-
and second-order integration algorithms, so we consider a finite
element space $V_h$ of continuous functions that are affine over each
element of the mesh. Discrete thermo-elastic deformations
$(\bm{\varphi}_h(\cdot,t),\Phi_h(\cdot,t))\in V_h^{d+1}$ 
can then be expressed as
\begin{equation}\label{SpaceDiscrete}
  (\bm{\varphi}_h({\bf X},t),\Phi_h({\bf X},t))=
  \Big(\sum_{a=1}^{N}\tn{N}_a(\Ne X)\bm\varphi_a(t),
  \sum_{a=1}^{N}\tn{N}_a(\Ne X)\Phi_a(t)\Big),
\end{equation}
  where $\bvarphi_a=(\varphi_a^1,...,\varphi_a^d)$ and $\Phi_a$
denote the value at node $a$ of the motion and the thermal position,
respectively, at any time. Correspondingly, $\{N_a\}_{a=1}^N$ is the
set of dual basis functions in $V_h$ to this choice of degrees of
freedom for each scalar field, i.e., the standard ``hat'' functions
over meshes of triangles or tetrahedra. 
%\footnote{Notation: the subscript $h$ is reserved for quantities depending on FE interpolations and lowercase latin letters $a$, $b, ...$ are used for referring to the nodal values of a given quantity.}. 

%and the vectors $\bm\Phi=(\Phi_1,...,\Phi_{n_d})\in\RU^{n_d}$ and $\bm\varphi=(\varphi_1^1,...,\varphi_{n_d}^d)\in\RU^{d\times n_d}$ are defined to denote the collection all the nodal values of mechanical and thermal positions, respectively.

We next obtain the semi-discrete equations that describe the evolution
in time of the degrees of freedom in the mesh.  To this end, we
approximate the values of
(\ref{KineticEnergy}), (\ref{InternalPotential1}) and
(\ref{ExternalPotential2})  by  selecting
quadrature rules, namely, we define functions $\textsf K_h\colon V_h^d \to
\mathbb R$, $\textsf W_h\colon V_h^d \times V_h \times V_h \to \mathbb
R$, and $\textsf E_h \colon V_h^d \times V_h \to \mathbb R$ as
\begin{subequations}
\begin{eqnarray}
   \tn{\textsf{K}}_h({\Ne v}_h)
   &=&
   \dfrac{\rho_0}{2}\sum_{i=1}^{N_q}w_i
   {\Ne v}_h(\bm\xi_i)\cdot
   {\Ne{ v}}_h(\bm\xi_i),%=\sum_{a=1}^{n_d}\tn m_a\dot{\bm\varphi}_a\cdot\dot{\bm\varphi}_a,
   \label{SemiDiscreteKinetic}
   \\
   \tn{\textsf W}_h(\bm\varphi_h,\Phi_h,\theta_h)
  &=&
  \rho_0\sum_{i=1}^{N_q}w_i
  \tn{\textsf{A}}\left(\nabla\bm\varphi_h(\bm\xi_i),
  \nabla\Phi_h(\bm\xi_i),\theta_h(\bm\xi_i)\right),
  \label{SemiDiscretePotetialU}
  \\
  \tn{\textsf E}_h(\bm\varphi_h,\Phi_h)
   &=&
  \sum_{i=1}^{N_q}w_i\rho_0\big(V_B(\bvarphi_h(\bm{\xi}_i))- \tn {\textsf{Q}}\;
  \Phi_h(\bm \xi_i)\big)
-
  \sum_{j=1}^{N_{t}}w_j^t (\Ne T \cdot \bvarphi)(\bm \xi^t_j) -
  \sum_{j=1}^{N_{h}}w_j^h
  (\overline{\tn h}\; \Phi)(\bm \xi_j^h),
  \label{SemiDiscretePotentialE}
\end{eqnarray}
\end{subequations}
where $\{w_i,\bm\xi_i\}_{i=1}^{N_q}$,
$\{w_j^h,\bm\xi^h_j\}_{j=1}^{N_h}$ and
$\{w_j^t,\bm\xi^t_j\}_{j=1}^{N_t}$ are quadrature points and
weights over $\Omega$, $\Gamma_h$ and $\Gamma_t$, respectively. The
Lagrangian for the semi-discrete system is
\begin{equation}
  \label{eq:2}
  \tn{\textsf L}_h(\bm\varphi_h,\Phi_h,{\Ne v}_h,\theta_h) =
  \tn{\textsf{K}}_h({\Ne v}_h) - \tn{\textsf W}_h(\bm\varphi_h,\Phi_h,\theta_h)- \tn{\textsf E}_h(\bm\varphi_h,\Phi_h).
\end{equation}
The equations of motion for the semi-discrete system follow from the
application of Hamilton's principle with this Lagrangian, as described next.
Thermo-elastic deformations $(\bvarphi_h,\Phi_h)$ of the
semi-discrete system can be succinctly specified through the evolution
of all nodal values
$(\bvarphi_a(\cdot),\Phi_a(\cdot))$, where unless otherwise specified, $a$ runs from $1$
to the
number of nodes in the mesh $N$. Consider then the set $\ca C_h$ of smooth enough motions
$(\bvarphi_a(\cdot),\Phi_a(\cdot))$ during the time interval $[0,T]$ between two
prescribed configurations
$(\bm\varphi_a(0),\Phi_a(0))=(\bm{\varphi}_{0}(\Ne
X_a),\Phi_{0}(\Ne X_a))$ and
$(\bm\varphi_a(T),\Phi_a(T))=(\bm{\varphi}_{T}(\Ne
X_a),\Phi_{T}(\Ne X_a))$,
which satisfy the boundary conditions
\begin{equation}\label{Dirichlet_1h}
  \begin{aligned}
 \bm{\varphi}_a(t)& =&\overline{
   \bm\varphi }(\Ne X_a,t), & \quad \tn{for all }\Ne X_a \in \Gamma_\varphi\\
 \Phi_a(t)&=&\overline{ \Phi }(\Ne X_a,t), & \quad \tn{for all }\Ne X_a \in \Gamma_\Phi,    
  \end{aligned}
 \end{equation} 
for all $t\in [0,T]$. Here $\Ne X_a\in \overline\Omega$ denotes the
position of node $a$. Trajectories of the system are the stationary
points in $\ca C_h$ of the action functional
\begin{equation}\label{SemiDiscreteActionFunctional}
  \tn{\textsf S}_h\left(\bm\varphi_a(\cdot),\Phi_a(\cdot)\right)=
  \int_0^T\tn{\textsf
    L}_h(\bm{\varphi}_h,\Phi_h,\dot{\bm{\varphi}}_h,\dot\Phi_h)\; dt.
\end{equation}
The corresponding Euler-Lagrange equations are
\begin{subequations}
\label{eq:5}
\begin{align}
  \sum_{b=1}^{N}\tn m_{ab}\ddot{\bm\varphi_b}
  &=-\Ne S_a+\Ne B_a,
  & \Ne X_a\not \in \Gamma_\varphi
  \label{SemiDiscreteBalance1}
  \\
  \frac{\tn d}{\tn d t}
  \Upsilon_a(\bm\varphi_h,\Phi_h,\dot\Phi_h)
  &=
  \tn{\textsf Q}_a+\tn H_a, 
  &\Ne X_a\not \in \Gamma_\Phi
  \label{SemiDiscreteBalance2}
\end{align}
for $a=1,\ldots,N$, 
where  
\begin{equation}
  \label{eq:4}
\tn m_{ab} = {\rho_0} \sum_{i=1}^{N_q} w_i \tn N_a(\bm \xi_i) \tn
N_b(\bm \xi_i)% \frac{\partial \textsf K_h}{\partial \dot \varphi_a\partial \dot \varphi_b}  
\end{equation}
is the $ab$-component of the consistent mass
matrix $m$, and
\begin{align}
  \label{eq:3}
  \Ne S_a & = \frac{\partial \tn{\textsf W}_h}{\partial{\bm\varphi_a}}, 
& \Ne B_a & = -\frac{\partial \tn{\textsf E}_h}{\partial{\bm\varphi_a}}, 
& \tn{\textsf Q}_a & = -\frac{\partial \tn{\textsf E}_h}{\partial\Phi_a},
& \Upsilon_a = -\frac{\partial \tn{\textsf W}_h}{\partial{\theta_a}},  & 
& \tn{\tn H}_a & = -\frac{\partial \tn{\textsf W}_h}{\partial\Phi_a}.
\end{align}
All partial derivatives in \eqref{eq:5} are
computed by keeping all other nodal values constant. These equations
have to be complemented with appropriate initial and boundary
conditions.
\end{subequations}

%A motion of the body is obtained by means of considering curves in $\ca Q_h$ parametrized in function of time. To build such curves consider the set $\ca C_h$ of smooth enough trajectories of the nodal values of the mechanical and thermal positions
%
%\begin{equation}\label{NodalValuesPositions}
%   (\bm\varphi_a(\cdot),\Phi_a(\cdot)):[0,T]\rightarrow\RU^{d+1},
%   \qquad a\in\{1,...,n_d\},
%\end{equation}
%
%between two prescribed configurations $\left\{\bm\varphi_a(0),\bm\varphi_a(0)\right\}_{a=1}^{n_d}$ and $\left\{\bm\varphi_a(T),\Phi_a(T)\right\}_{a=1}^{n_d}$ which are compatible with the boundary conditions detailed in (\ref{Dirichlet_1}).  The semi-discrete (discrete in space) trajectories of material points $\bm\xi\in K_e\subset\ca T_h$ are obtained considering the FE interpolation given (\ref{SpaceDiscrete}) thus yielding to the trajectories $(\bm\varphi_h(\bm\xi,\cdot),\Phi_h(\bm\xi,\cdot))$. 

The semi-discrete problem \eqref{eq:5}, obtained after introducing the
finite element discretization in space, is a system of ordinary
differential equations (ODEs) for $(\bm \varphi_a(\cdot),
\Phi_a(\cdot))$. Therefore, most of the standard techniques for
solving ODEs numerically apply, including in this case structured or
geometric integrators, e.g., \cite{Hairer1}. Of course, a new
semi-discrete problem is obtained for each mesh, but the $L^2$-norm of
$(\bm\varphi-\bm\varphi_h,\Phi-\Phi_h, \dot {\bm \varphi}-\dot{\bm
  \varphi}_h,\theta-\theta_h)$ is expected to be $\ca O(h^2)$ for
smooth solutions. We show this with examples later on.

% Apart from that, other sources of errors exist\footnote{See
%   \cite{CiarletFEM} for an account about error analysis in FE
%   discretizations of differential equations.}: (i) the error due to
% the inexact evaluation of the volume integrals in
% (\ref{SemiDiscreteKinetic}), (\ref{SemiDiscretePotetialU}) and
% (\ref{SemiDiscretePotentialE}) and (ii) the error due to the inexact
% approximation of the boundary conditions (\ref{Dirichlet_1}) which are
% exactly enforced only at the nodes as it is shown in
% (\ref{Dirichlet_1h}). In particular, the selection of the quadrature
% rule has important effects on the rate of convergence of the FE method
% and its stability in dynamic simulations
% \cite{Durufle01,Bathe1,Cohen001

%The corresponding semi-discretization of the action functional is given by
%
%were the dependence on the trajectories of the nodal variables has been highlighted. The application of Hamilton's principle yields to the following system of ODEs corresponding to the semi-discrete E-L equations,
%

\paragraph{Hamiltonian perspective} The transition to the Hamiltonian
point of view may be carried out using the Legendre transform to
compute the conjugate values of the mechanical and thermal momentum as
\begin{subequations}
\begin{equation}\label{BCPositionMomentumForm}
  \Ne p_a=
  \frac{
  \partial\tn{\textsf L}_h
  }
  {
  \partial\dot{\bm\varphi}_a
  }=
  \sum_{b=1}^{n_d}\tn m_{ab}\dot{\bm\varphi}_b
  \qquad
  \tn{and}
  \qquad
  \tau_a=
  \frac{
  \partial
  \tn{\textsf L}_h
  }
  {
  \partial\theta_a
  } = -
  \frac{
  \partial
  \tn{\textsf W}_h
  }
  {
    \partial\theta_a
  } = \Upsilon_a(\bm\varphi_h,\Phi_h, \theta_h),
\end{equation}
respectively, for any node $a$, where again partial derivatives are
computed keeping all other nodal values constant. In the following we
will use
\begin{equation}
  \label{eq:13}
  \Ne p = \{\Ne p_1, \ldots,\Ne p_N\} \quad \text{ and } \quad \bm
  \tau = \{ \tau_1,\ldots,\tau_N\}.
\end{equation}
 The function $\Ne p_a(\cdot)$ is invertible,
since the mass matrix is. If for each set of values of $(\bm
\varphi_h,  \Phi_h)$ the discrete Lagrangian
$\tn{\textsf L}_h$ is a strictly convex function of $\theta_h$, then the
function $\Upsilon_a(\bm \varphi_h, \Phi_h,\cdot)$ is
invertible. This is guaranteed because we assumed that $\textsf A$ is
a strictly convex function of
$\theta$, the same requirement needed for the definition of $\hat
\theta$ in \eqref{Temperature}. Under these conditions, we denote this
inverse with 
\begin{equation}
  \label{eq:10}
  \hat \theta_h(\Ne X; \bm \varphi_h, \Phi_h, \bm \tau) =
  \sum_{a=1}^d  \tn N_a(\Ne X) \hat\theta_a (\bm \varphi_h, \Phi_h,
  \bm \tau) \in V_h.
\end{equation}
where $\{\hat \theta_a\}_a$ are its nodal values. Notice that $\hat \theta_h$ is a different function than $\hat \theta$
in \eqref{Temperature}. Correspondingly, we define the discrete
entropy density field as
\begin{equation}
  \label{eq:12a}
  \eta_h (\Ne X; \bm \varphi_h, \Phi_h, \bm \tau) =
  \eta(\nabla \bm\varphi_h(\Ne X),
  \hat \theta_h(\Ne X; \bm \varphi_h, \Phi_h,\bm \tau)),
\end{equation}
where, although not essential, we have used the assumed independence
of $\eta$ on $\bm \beta$. 
The Hamiltonian for the semi-discrete system then follows as
\begin{align}
  \tn{\textsf H}_h(\bm\varphi_h,&\Phi_h,\Ne p, \bm \tau)  =
  \sum_{b=1}^N \left(\tau_b \hat \theta_b + \Ne p_b \sum_{a=1}^N (m^{-1})_{ba} \Ne p_a\right) -
  \textsf L_h \label{eq:28}\\
  & =  \dfrac{1}{2}\sum_{a,b=1}^{n_d}\Ne p_a\cdot\tn (m^{-1})_{ab}\Ne p_b+\rho_0
  \sum_{p=1}^{N_q}w_p\tn{\textsf
    U}(\nabla
  \bm\varphi_h(\bm\xi_p),\nabla\Phi_h(\bm\xi_p),\eta_h(\bm\xi_p;\bm
  \varphi_h, \Phi_h, \bm \tau))+
  \tn{\textsf E}_h(\bm\varphi_h,\Phi_h).    \nonumber
\end{align}
where $\tn{\textsf U}$ is the internal energy density per unit mass
from \eqref{eq:8}.  
%The corresponding symplectic form  is given by
%
%\begin{equation}
%  \bm\omega_h
%   =
%   \sum_{a=1}^{n_d}
%   \left(\sum_{i=1}^{d}
%   \varphi_a^{i}\wedge\tn p_a^{i}
%   +
%   \Phi_a\wedge\tau_a
%   \right).
% \end{equation}
% %
 \end{subequations}
%In the appendix we provide the Hamiltonian for the continuous problem

\comment{
This is how we obtain the Hamiltonian.
\begin{equation}
  \label{eq:10}
  \begin{aligned}
  \tau_a  & = -\frac{\partial \textsf W_h}{\partial \theta_a}(\bm
  \varphi_h, \Phi_h, \theta_h)
   = \rho_0 \sum_{i=1}^{N_q} w_i \eta(\nabla \bm\varphi_h(\bm \xi_i),
  \nabla \Phi_h(\bm \xi_i),  \theta_h(\bm \xi_i)) \tn N_a(\bm \xi_i)
  \end{aligned}
\end{equation}
So,
\begin{equation}
  \label{eq:12b}
  \begin{aligned}
  \sum_{b=1}^N \tau_b &\theta_b + \textsf W_h(\bm \varphi_h,\Phi_h, 
  \theta_h)   =\\
&= \rho_0 \sum_{i=1}^{N_q} w_i \left[ \eta(\nabla \bm\varphi_h(\bm \xi_i),
  \nabla \Phi_h(\bm \xi_i), \theta_h(\bm \xi_i))\ \sum_{b=1}^N \tn
  N_b(\bm \xi_i) \theta_b + \textsf A(\nabla \bm\varphi_h(\bm \xi_i),
  \nabla \Phi_h(\bm \xi_i), \theta_h(\bm \xi_i))\right]\\
& = \rho_0 \sum_{i=1}^{N_q} w_i \left[ \eta(\nabla \bm\varphi_h(\bm \xi_i),
  \nabla \Phi_h(\bm \xi_i), \theta_h(\bm \xi_i))\   \theta_h(\bm \xi_i) + \textsf A(\nabla \bm\varphi_h(\bm \xi_i),
  \nabla \Phi_h(\bm \xi_i), \theta_h(\bm \xi_i))\right]\\
  &= \rho_0 \sum_{i=1}^{N_q} w_i \textsf U(\nabla \bm\varphi_h(\bm \xi_i),
  \nabla \Phi_h(\bm \xi_i), \eta(\nabla \bm\varphi_h(\bm \xi_i),
  \nabla \Phi_h(\bm \xi_i), \theta_h(\bm \xi_i))) 
  \end{aligned}
\end{equation}
}
%
%-----------------------------------------------------------------
%
\subsection{Discretization in time: variational integrators}\label{sec:discr-time:-vari}
The final step consists in constructing  approximations in time to the
semi-discrete trajectories that solve \eqref{eq:5}. To this end, we
partition the time interval $[0,T]$  with a time-step $\Delta
t=T/M$,
$M>1$, and  set $t^k=k T/M$, $k=0,\ldots,M$.  We denote by
$(\bm\varphi_a^k,\Phi_a^k)$  the approximation of
$(\bm\varphi_a(t^k),\Phi_a(t^k))$ computed by the algorithm, and use
(\ref{SpaceDiscrete}) to compute $(\bm\varphi_h(\Ne X,t^k),\Phi_h(\Ne
X,t^k))$ as the approximation to $(\bm\varphi(\Ne X,t^k),\Phi(\Ne
X,t^k))$  for all $\Ne X\in K, \; \forall K\subset\ca T_h$.

The construction of variational integrators proceeds by defining
approximations of the action obtained from \eqref{eq:2}. Since in this
case the Lagrangian depends on the velocities in an other than
quadratic way  (specifically, the temperature), some care is needed to
construct explicit second-order integrators in time. 
%For simplicity,
%we assume that the {\it functions $\overline{\bm\varphi}(\Ne X_a,t)$ and
%$\overline \Phi(\Ne X_a,t)$ in \eqref{Dirichlet_1h} are affine
%respect to time}. We detail the more general case in \S \ref{sec:second-order-with}.
In the following
we follow the ideas in \cite{MataLew1}. We begin by defining the two
discrete Lagrangians
\begin{equation}
  \label{eq:6}
  \begin{aligned}
    {\textsf L}_{h,\Delta t}^i(\bm\varphi_h^0,\Phi_h^0,\bm
    \varphi_h^1,\Phi_h^1) &
    = \Delta t\; \textsf L_h\left(\bm \varphi_h^i,\Phi_h^i,\frac{\bm
      \varphi_h^1-\bm\varphi_h^0}{\Delta t},
    \frac{\Phi_h^1-\Phi_h^0}{\Delta t}\right)
% \\
%     {\textsf L}_{h,\Delta t}^1(\bm\varphi_h^0,\Phi_h^0,\bm
%     \varphi_h^1,\Phi_h^1) &
%     = \Delta t\; \textsf L_h\left(\bm \varphi_h^1,\Phi_h^1,\frac{\bm
%       \varphi_h^1-\bm\varphi_h^0}{\Delta t},
%     \frac{\Phi_h^1-\Phi_h^0}{\Delta t}\right),
  \end{aligned}
\end{equation}
for $i=0,1$, and from here the class of discrete Lagrangians
\begin{equation}
  \label{eq:7}
\textsf  L_{h,\Delta t}^\alpha = (1-\alpha) \textsf L_{h,\Delta t}^0+ \alpha
 \textsf  L_{h,\Delta t}^1
\end{equation}
for any $\alpha\in [0,1]$. For completeness, the expressions for the
discrete Lagrangians in \eqref{eq:6} are
\begin{subequations}
\begin{eqnarray}
  \tn{\textsf{L}}^{i}_{h,\Delta t}
  &=&
  \Delta t\left(
    \sum_{a,b=1}^N \frac{m_{ab}}{2\Delta t^2} (\bm\varphi_a^1-\bm\varphi_a^0)\cdot(\bm\varphi_b^1-\bm\varphi_b^0)-
  \tn{\textsf W}_{h}\left(\bm\varphi_h^i,\Phi_h^i,\frac{\Phi_h^1-\Phi_h^0}{\Delta t}\right)-
  \tn{\textsf E}_{h}(\bm\varphi_h^i,\Phi_h^i)\right),
  % \\
  % \tn{\textsf{L}}^{1}_{h,\Delta t}
  % &=&
  % \Delta t\left(
  % \tn{\textsf K}_{h,\Delta t}(\bm\varphi_h^{1},\bm\varphi_h^0)-
  % \tn{\textsf W}^{1}_{h,\Delta t}(\bm\varphi_h^1,\Phi_h^0,\Phi_h^1)-
  % \tn{\textsf E}_{h}(\bm\varphi_h^{1},\Phi_h^{1})
  % \right),
\end{eqnarray}
\end{subequations}
for $i=0,1$.
% , where
% \begin{equation}
%   \tn{\textsf K}_{h,\Delta t}(\Delta \bm\varphi_h)=
%   \dfrac{1}{2\Delta t^2}\sum_{a,b=1}^{N}\tn m_{ab}
%   \Delta\bm\varphi_b\cdot
%   \Delta\bm\varphi_a,
% \end{equation}
% and
% \begin{eqnarray}
%   \tn{\textsf W}_{h}(\bm\varphi_h,\Phi_h,\Delta\Phi_h)
%   &=&
%   \rho_0\sum_{j=1}^{N_q}w_j\tn{\textsf{A}}
%   \left(
%   \nabla\bm{\varphi}_h(\bm{\xi}_j),
%   \nabla\Phi_h(\bm{\xi}_j),
%   \Delta\Phi_h(\bm\xi_j)
%   \right).
%   \label{PotentialDiscrete1}
%   % \\
%   % \tn{\textsf W}_{h,\Delta t}^{1}(\bm\varphi_h,\Phi_h^{0},\Phi_h^1)
%   % &=&
%   % \rho_0\sum_{i=1}^{N_q}w_i\tn{\textsf{A}}
%   % \left(
%   % \nabla\bm\varphi_h(\bm\xi_i),
%   % \nabla\Phi_h^1(\bm\xi_i),
%   % \frac{\Phi_h^{1}(\bm\xi_i)-\Phi_h^{0}(\bm\xi_i)}{\Delta t}
%   % \right).  
%   % \label{PotentialDiscrete2}  
% \end{eqnarray}
For each $\alpha$, the algorithm follows by forming the discrete
action sum 
\begin{equation}
  \label{eq:8b}
 {\sf S}_d(\bm \varphi_h^0,\Phi_h^0, \ldots, \bm \varphi_h^M,\Phi_h^M) =
  \sum_{k=0}^{M-1} {\sf L}_{h, \Delta t}^\alpha(\bm \varphi_h^k, \Phi_h^k,\bm \varphi_h^{k+1},\Phi_h^{k+1})
\end{equation}
and finding $(\bm \varphi_h^0,\Phi_h^0, \ldots,
\bm\varphi_h^M,\Phi_h^M)$ satisfying the boundary conditions
\eqref{Dirichlet_1h} at each time $t^k$, $k=0,\ldots, M$, 
that is a
stationary point of $\sf S_d$ among all admissible variations. These are
the variations that keep $(\bm\varphi_h^0,\Phi_h^0)$ and ($\bm
\varphi_h^N,\Phi_h^N)$ fixed and are zero on the Dirichlet boundary of
each field $(\Gamma_\varphi$ and $\Gamma_\Phi$).
This leads to the Discrete
Euler-Lagrange equations, which in position-momentum form (see
\cite{Marsden2}) read
%\begin{subequations}
% \begin{eqnarray}
%   \Ne p_a^k 
% %  &=&
% %  \hat{\Ne p}\left(...,\bm\varphi_j^{k+1},\Phi_j^{k+1},...\right),
% %  \nonumber
% %  \\
%   &=& - D_1 \textsf L_{h,\Delta t}^\alpha\left(
%   \bm\varphi_h^k,\Phi_h^k,\bm\varphi_h^{k+1},\Phi_h^{k+1}\right)=
%   \sum_{b=1}^{N}\tn m_{ab}
%   \frac{(\bm\varphi_b^{k+1}-\bm\varphi_b^k)}{\Delta t}
%   +(1-\alpha)\Delta t
%   \left(
%   \Ne S^{k+}_a-\Ne J_a^k
%   \right),
%   \label{Eq1} 
%   \\
%   \Ne p_a^{k+1}
%   &=& \phantom{-}D_3 \textsf L_{h,\Delta t}^\alpha\left(
%   \bm\varphi_h^k,\Phi_h^k,\bm\varphi_h^{k+1},\Phi_h^{k+1}\right)=
%   \sum_{b=1}^{N}\tn m_{ab}
%   \frac{(\bm\varphi_b^{k+1}-\bm\varphi_b^k)}{\Delta t}
%   -\alpha\Delta t
%   \left(
%   \Ne S^{k+1-}_a-\Ne J_a^{k+1}
%   \right),  
%   \label{Eq3}
%   \\
% %\end{eqnarray}
% %\begin{eqnarray}
%   \tau_a^k
% %  &=&
% %  \hat{\tau}(...,\bm\varphi_j^{k+1},\Phi_j^{k+1},...),
% %  \nonumber
% %  \\
%   &=& -D_2 \textsf L_{h,\Delta t}^\alpha\left(
%   \bm\varphi_h^k,\Phi_h^k,\bm\varphi_h^{k+1},\Phi_h^{k+1}\right)=
%   (1-\alpha)\Upsilon_a^{k+}+\alpha\Upsilon_a^{k+1-}-
%   (1-\alpha)\Delta t
%   \left(
%   \tn H_a^k+\tn R_a^k
%   \right),   
%   \label{Eq2}
%   \\
%   \tau^{k+1}_a
%   &=& \phantom{-}D_4 \textsf L_{h,\Delta t}^\alpha\left(
%   \bm\varphi_h^k,\Phi_h^k,\bm\varphi_h^{k+1},\Phi_h^{k+1}\right)=
%   (1-\alpha)\Upsilon_a^{k+}+\alpha\Upsilon_a^{k+1-}+
%   \alpha\Delta t
%   \left(
%   \tn H_a^{k+1}+\tn R_a^{k+1}
%   \right),  
%   \label{Eq4}   
% \end{eqnarray}
% \end{subequations}
%
%
\begin{subequations}
\label{eq:9}
\begin{eqnarray}
  \Ne p_a^k 
%  &=&
%  \hat{\Ne p}\left(...,\bm\varphi_j^{k+1},\Phi_j^{k+1},...\right),
%  \nonumber
%  \\
  &=& - D_{1a} \textsf L_{h,\Delta t}^\alpha\left(
  \bm\varphi_h^k,\Phi_h^k,\bm\varphi_h^{k+1},\Phi_h^{k+1}\right)=
  \sum_{b=1}^{N}\tn m_{ab}
  \frac{(\bm\varphi_b^{k+1}-\bm\varphi_b^k)}{\Delta t}
  +(1-\alpha)\Delta t
  \left(
  \Ne S^{0,k}_a-\Ne B_a^k
  \right),
  \label{Eq1} 
  \\
  \Ne p_a^{k+1}
  &=& \phantom{-}D_{3a} \textsf L_{h,\Delta t}^\alpha\left(
  \bm\varphi_h^k,\Phi_h^k,\bm\varphi_h^{k+1},\Phi_h^{k+1}\right)=
  \sum_{b=1}^{N}\tn m_{ab}
  \frac{(\bm\varphi_b^{k+1}-\bm\varphi_b^k)}{\Delta t}
  -\alpha\Delta t
  \left(
  \Ne S^{1,k}_a-\Ne B_a^{k+1}
  \right),  
  \label{Eq3}
  \\
%\end{eqnarray}
%\begin{eqnarray}
  \tau_a^k
%  &=&
%  \hat{\tau}(...,\bm\varphi_j^{k+1},\Phi_j^{k+1},...),
%  \nonumber
%  \\
  &=& -D_{2a} \textsf L_{h,\Delta t}^\alpha\left(
  \bm\varphi_h^k,\Phi_h^k,\bm\varphi_h^{k+1},\Phi_h^{k+1}\right)=
  (1-\alpha) \Upsilon_a^{0,k} + \alpha
  \Upsilon_a^{1,k}-(1-\alpha) \Delta t\;\left( \tn H_a^{0,k}+ \textsf Q_a^k\right),
  \label{Eq2}
  \\
  \tau^{k+1}_a
  &=& \phantom{-}D_{4a} \textsf L_{h,\Delta t}^\alpha\left(
  \bm\varphi_h^k,\Phi_h^k,\bm\varphi_h^{k+1},\Phi_h^{k+1}\right)=
  (1-\alpha)\Upsilon_a^{0,k}+\alpha\Upsilon_a^{1,k}+
  \alpha\Delta t
  \left(
   \tn H_a^{1,k} + \textsf Q_a^{k+1}
  \right),  
  \label{Eq4}   
\end{eqnarray}
\end{subequations}
for each nodal value admitting a variation, where $D_{ia}\textsf L_{h,\Delta t}^\alpha$ denotes the partial
derivative of $\textsf L_{h,\Delta t}^\alpha$ with respect to the
value at node $a$ of its
$i$-th argument, $\Ne{p}^k_a$ is numerical approximation to the nodal
value of the mechanical momentum at $t^k$, and $\tau^k_a$ is the
corresponding approximation to the thermal momentum. Additionally,
\begin{equation}
  \label{eq:6b}
  \begin{aligned}
 \Ne S_a^{i,k} =
%\frac{\partial \textsf W_{h}}{\partial \bm \varphi_a}
\Ne S_a
\left(\bm\varphi_h^{k+i},\Phi_h^{k+i},\frac{\Phi_h^{k+1}-\Phi_h^k}{\Delta
  t}\right), \qquad 
\Ne B_a^k = \Ne B_a
%-\frac{\partial \textsf E_h}{\partial  \bm\varphi_a}
(\bm\varphi_h^k,\Phi_h^k),\qquad
\textsf
Q_a^k=\textsf Q_a
%-\frac{\partial \textsf E_h}{\partial \Phi_a}
(\bm \varphi_h^k,\Phi_h^k),
\\
  \Upsilon_a^{i,k} = \Upsilon_a
%-\frac{\partial\textsf W_h}{\partial  \theta_a}\
\left(\bm\varphi^{k+i}_h,\Phi_h^{k+i},\frac{\Phi_h^{k+1}-\Phi_h^k}{\Delta
    t}\right), \qquad 
\tn H_a^{i,k} =\tn H_a
%-\frac{\partial \textsf W_{h}}{\partial \Phi_a} 
\left(\bm
\varphi_h^{k+i},\Phi_h^{k+i},\frac{\Phi_h^{k+1}-\Phi_h^k}{\Delta t}\right),     
  \end{aligned}
\end{equation}
for $i=0,1$.
Explicit
expressions for $\Ne S_a^{i,k}$, $\Upsilon_a^{i,k}$, $H_a^{i,k}$,
$\Ne B_a^k$, ${\textsf Q}_a^k$ are given in  \ref{sec:expl-expr-deriv}.

%
% To this end,
% we approximate the time derivatives of $\bm\varphi_a(t)$ and
% $\Phi_a(t)$ in  each time interval $(t^k,t^{k+1})$, $k\ge
% 0$, as
% %
% \begin{equation}\label{velocity}
%   \dot{\bm{\varphi}}_a(t)
%   \approx
%   \frac{\bm{\varphi}_a^{k+1}-\bm{\varphi}_a^k}{\Delta t}
%   \quad\tn{and}\quad
%   \quad\theta_a(t)
%   \approx
%   \frac{\Phi_a^{k+1}-\Phi_a^k}{\Delta t},
%   \qquad
% \end{equation}
% for $t\in(t^k,t^{k+1})$.%
%
%\paragraph{Remark} The above aproximations used for temperature and velocity are  discontinuous across time interval boundaries as explained in \cite{MataLew1}. However, they are consisten with assuming that for fixed $\Ne z\in K_e$ the time continuous trajectories $(\bm\varphi_h(\Ne z,\cdot),\Phi_h(\Ne z,\cdot))$ are linear piecewise continuous polynomials in $[t^0,t^1]$ which joins $(\bm\varphi_h^0(\Ne z),\Phi_h^0(\Ne z))$ and $(\bm\varphi_h^1(\Ne z),\Phi_h^1(\Ne z))$. Moreover, VIs of arbitrarily high order may be developed for finite dimensional thermoelastic problems by means of choosing higher order polynomials\cite{MataLew1,Lew1,Marsden2} $\blacksquare$
%
%
%
%--------------------------------------------------------------------------------------------------------------
%

\paragraph{Algorithm} Equation \eqref{eq:9} defines an algorithm,
which given $(\bm \varphi_h^k, \Phi_h^k, \Ne p^k, \bm \tau^k)$, returns $(\bm
\varphi_h^{k+1},\allowbreak  \Phi_h^{k+1}, \Ne p^{k+1},\bm \tau^{k+1})$, for any $k\ge 0$.  The basic procedure is as follows:
\begin{enumerate}
\item[(i)] Solve  (\ref{Eq1}) and (\ref{Eq2}) for $(\bm\varphi_h^{k+1},\Phi_h^{k+1})$. 
\item[(ii)] Update $(\Ne p^{k+1},\tau^{k+1})$ with  (\ref{Eq3})
  and (\ref{Eq4}), for $a=1,\ldots,N$. 
\item[(iii)] If of interest, an approximation of the temperature may be computed as
  \begin{equation}
    \label{eq:12}
    \theta_h^{k+1}(\Ne X) = \hat \theta_h(\Ne X; \bm \varphi_h^{k+1},
    \Phi_h^{k+1}, \bm \tau^{k+1}),
  \end{equation}
and that of the velocity as
\begin{equation}
  \label{eq:14}
  \dot{\bm \varphi}^{k+1}_h(\Ne X) = \sum_{a=1}^N \sum_{b=1}^N (m^{-1})_{ab}
  \Ne p_b^{k+1} \tn N_a(\Ne X).
\end{equation}
\end{enumerate}

If a Gauss-Legendre (optimal) quadrature rule
is used to evaluate the volume integrals, the resulting algorithm is
implicit due to the coupling between different degrees of freedom
introduced by the consistent mass matrix, by $\tn
H_a^{i,k}$, and by $\Upsilon_a^{i,k}$. 

%
%--------------------------------------------------------------------------------------------
%
\section{Explicit algorithms}\label{ExplicitSecondOrder}

In this section we derive two explicit schemes that exhibit
second-order accuracy in both $h$ and $\Delta t$ by combining two
ingredients: (i) a Gauss-Lobatto (GL) quadrature rule to compute the
volume integrals appearing in (\ref{SemiDiscreteKinetic}),
\eqref{SemiDiscretePotetialU}, and \eqref{SemiDiscretePotentialE},
and (ii) a half-step composition of the methods obtained for
$\tn{\textsf L}_{h,\Delta t}^0$ and $\tn{\textsf L}_{h,\Delta
  t}^1$. 

The algorithms below are explicit in the sense that there is no
coupled system of equations that needs to be solved. However, a
nonlinear solver for a single scalar value could be needed at each
node, depending on the constitutive relation.

\subsection{Gauss-Lobatto quadrature rule}
The GL rule adopts integration points at the nodes of the
elements, which notably eliminates the coupling among different
degrees of freedom.  In particular, a lumped mass matrix is
obtained. 
To compute the values of \eqref{eq:6b}  under these conditions
 some care is needed, since $\nabla \tn N_a$ is
discontinuous at nodes. An explicit evaluation of the expressions
given in
\ref{sec:expl-expr-deriv} is not possible. Instead, the appropriate limit as the
quadrature points approach each node from within each element is
used. To make the resulting expressions explicit, it is convenient to
define
\begin{equation}
  \label{eq:15}
  w_{a,K} = \frac{1}{d+1} \text{Volume}(K)
\end{equation}
for each node $a$ and each element $K$ in the ring of $a$, namely,
\begin{equation}
  \label{eq:17}
  \text{ring}(a) = \left\{K\in \ca T_h\mid a\in K\right\}.
\end{equation}
 The weight
of the quadrature point at $a$ is then
\begin{equation}
  \label{eq:16}
    w_a = \sum_{\text{ring}(a)} w_{a,K},
\end{equation}
and the lumped mass matrix \eqref{eq:4} becomes $ \tn m_{ab} = \rho_0
\delta_{ab} w_a$(no sum over $a$). With these definitions,  and noticing that
$\bm\xi_a=\Ne X_a$ and $\tn
N_a(\bm \xi_b)=\delta_{ab}$, we get
\begin{subequations}
\label{eq:19}
\begin{align}
  \Ne  S_a^{i,k} &=\Ne
  S_a\left(\bm\varphi_h^{k+i},\Phi_h^{k+i},\frac{\Phi_a^{k+1}-\Phi_a^k}{\Delta
    t}\right) = \sum_{\text{ring}(a)}  w_{a,K} \Ne P \left(\nabla \bvarphi_h^{k+i}|_K
, \nabla \Phi_h^{k+i}|_K,
\frac{\Phi_a^{k+1}-\Phi_a^k}{\Delta t}\right) \nabla \tn
N_a|_K \label{eq:ScompGL}\\
\Ne B_a^k&=\Ne B_a(\bm\varphi_h^{k})  =
-w_a\rho_0\frac{\partial V_B}{\partial
  \bvarphi}(\bvarphi_a^{k}) 
+
  \sum_{j=1}^{N_{t}}w_j^t \Ne T\; \tn N_a(\bm \xi^t_j) \label{eq:BcompGL}
\\
\textsf
Q_a^k &= \textsf Q_a=  w_a\rho_0 \tn {\textsf{Q}}\;
+  \sum_{j=1}^{N_{h}}w_j^h
  (\overline{\tn h}\; \tn N_a)(\bm \xi_j^h), \label{eq:QcompGL}
\\
  \Upsilon_a^{i,k} &=\Upsilon_a \left(\bm\varphi_h^{k+i},\frac{\Phi_a^{k+1}-\Phi_a^k}{\Delta
    t}\right) = \sum_{\text{ring}(a)} \rho_0  w_{a,K}
  \eta\left(\nabla\bm\varphi_h^{k+i}|_K,
    %\nabla\Phi_h^{k+i}|_K,
    \frac{\Phi_a^{k+1}-\Phi_a^k}{\Delta
    t}\right),\label{eq:UpsiloncompGL}\\
\tn H_a^{i,k} &= \tn H_a(\bm\varphi_h^{k+i},\Phi_h^{k+i}) =  \sum_{\text{ring}(a)} w_{a,K} \Ne h\left(\nabla\bm\varphi_h^{k+i}|_K,
  \nabla\Phi_h^{k+i}|_K
  %,\frac{\Phi_a^{k+1}-\Phi_a^k}{\Delta t}
\right) \nabla\tn N_a|_K,\label{eq:HcompGL}
\end{align}
\end{subequations}
where we have used the assumption that $\eta$ does not depend on $\bm
\beta$ and that $\Ne h$ does not depend on $\theta$, as well as the
form of $\textsf E_h$, to only keep the arguments that the functions
above depend on.  

Finally, as mentioned earlier, the nodal values of the temperature
$\theta_a^{k+1}$ needed for the temperature field $\theta_h^{k+1}$
in \eqref{eq:12} follow from solving \eqref{eq:10} for each node
$a$. With this choice of quadrature the equation to be solved for
$\theta_a^{k+1}$ takes
the form
\begin{equation}
  \label{eq:18}
  \tau_a^{k+1} = \Upsilon_a(\bm\varphi_h^{k+1}, \theta_a^{k+1})=\rho_0 \sum_{\text{ring}(a)} w_{a,K}
  \eta\left(\nabla\bm\varphi_h^{k+1}|_K,
    %\nabla\Phi_h^{k+i}|_K,
    \theta_a^{k+1}\right).
\end{equation}
For the two constitutive models in \S \ref{Example1Continuum} explicit
expressions for $\theta_a^{k+1}$ can be obtained, so no numerical
solutions are needed.

\subsection{First order with $\Delta t$}
\label{sec:first-order-with}
First, we consider the following numerical algorithms:
\begin{enumerate}
  %
  %-----------------------------------------------
  %
\item For $\tn{\textsf L}_{h,\Delta t}^0$ we obtain a method $ F_{h,\Delta t}^0(\bm \varphi_h^k, \Phi_h^k, \Ne p^k, \bm
  \tau^k)=\tn
 (\bm
\varphi_h^{k+1},\allowbreak  \Phi_h^{k+1}, \Ne p^{k+1},\bm \tau^{k+1})$ which consists of
  \begin{enumerate}
  \item Computing $\Phi_a^{k+1}$ from (\ref{Eq2}), \eqref{eq:QcompGL},
    \eqref{eq:UpsiloncompGL}, and \eqref{eq:HcompGL}. This step might
    require solving an equation similar to \eqref{eq:18} for each
    node, depending on the constitutive relation. \label{item:1}

    \item Computing $\bm \varphi_a^{k+1}$ from (\ref{Eq1}),
      taking advantage of the lumped mass matrix, together with
      \eqref{eq:ScompGL} and \eqref{eq:BcompGL}.
    
    \item Computing $(\Ne p_a^{k+1},\tau_a^{k+1})$ from (\ref{Eq3})
      and (\ref{Eq4}), together with  \eqref{eq:19}.
   \end{enumerate}
   
   This algorithm can be identified with a version of the {\it symplectic-Euler-A} method \cite{Leimkuhler001,Hairer1,HairerEulerSymplectic,Marsden2} formulated for the (non-separable) Hamiltonian appearing in the G-N-II theory.  
   %
   %-----------------------------------
   %
   \vskip .25cm
\item For $\tn{\textsf L}_{h,\Delta t}^1$ we obtain a method $ F_{h,\Delta t}^1(\bm \varphi_h^k, \Phi_h^k, \Ne p^k, \bm
  \tau^k)=\tn
 (\bm
\varphi_h^{k+1},\allowbreak  \Phi_h^{k+1}, \Ne p^{k+1},\bm \tau^{k+1})$ which consists of
  \begin{enumerate} 

    \item Computing $\bm \varphi_a^{k+1}$ from (\ref{Eq1}),
      taking advantage of the lumped mass matrix. 
         
    \item Computing $\Phi_a^{k+1}$ fron (\ref{Eq2}) and
      (\ref{eq:UpsiloncompGL}). This step might
    require solving an equation similar to \eqref{eq:18} for each
    node, depending on the constitutive relation. \label{item:2}

    \item Computing $(\Ne p^{k+1}_a,\tau^{k+1}_a)$ from (\ref{Eq3})
      and (\ref{Eq4}), together with \eqref{eq:19}.
   \end{enumerate}
   
   In this case, this algorithm can be identified with the {\it
     symplectic-Euler-B} method
   \cite{Marsden2,Leimkuhler001,Hairer1,HairerEulerSymplectic}.  
 \end{enumerate}
For the constitutive models in \S \ref{Example1Continuum} it is
possible to solve the equations in steps \ref{item:1} and \ref{item:2}
explicitly. For more general models, the strict convexity of $\textsf A$
with respect to $\theta$ 
implies the convexity of $\textsf W_h$ with respect to $\theta_h$ as well,
and from there the solvability of the resulting equations.

%
%----------------------------------------------------------------------------------------------------
%
\subsection{Second order with $\Delta t$ --- Autonomous case}
\label{SecondOrderDeltaT}

The methods in \S~\ref{sec:first-order-with} are explicit. However,
they are only first-order accurate with $\Delta t$ (see, e.g.,
\cite[pp. 402]{Marsden2},\cite{MataLew1}). A second-order algorithm
can be easily obtained by composing a first-order integrator with its
adjoint to produce a self-adjoint method, see
\cite[Ch. 2]{Hairer1}. Those methods are symmetric and therefore 
have an even-order of accuracy. When the Dirichlet boundary conditions
are constant in time ($\dot{\overline{\bm \varphi}}=\dot{\overline{\Phi}}=0$), the resulting mechanical system is governed by
an autonomous Lagrangian, and the adjoint  method to
$\tn{F}_{h,\Delta t}^{1}$ is $\tn{F}_{h,\Delta t}^{\ast
  1}=\tn{F}_{h,\Delta t}^{0}$ \cite{Marsden2,MataLew1}. We took
advantage of this fact to formulate second-order methods for discrete adiabatic
thermoelastic systems in  \cite{MataLew1}.
 Based on those ideas, we formulate a
 second-order accurate method $\hat{\tn F}_{h,\Delta
  t}^{10}$ as
\begin{equation}\label{CompositionGeneral}
  \hat{\tn{F}}_{h,\Delta t}^{10}=
  \tn{F}_{h,\Delta t/2}^1\circ
  \tn{F}_{h,\Delta t/2}^{\ast 1}=\tn{F}_{h,\Delta t/2}^1\circ
  \tn{F}_{h,\Delta t/2}^{0}.
\end{equation}
A straightforward computation shows that the above composition method corresponds to the variational integrator derived from the following discrete Lagrangian

% \begin{align}
% \label{eq:11}
%   \tn{\textsf{L}}_{h,\Delta
%     t}^{01}&(\bm\varphi_h^0,\Phi_h^0,\bm\varphi_h^1,\Phi_h^1; t)
%   =\\
% &  \underset{\substack{
%       \bm\varphi_a^{1/2} = \overline{\bm \varphi}(\bm
%       X_a,t)\text{ if } \bm X_a\in \Gamma_\varphi\\
%       \Phi_a^{1/2} = \overline{\Phi}(\bm
%       X_a,t)\text{ if } \bm X_a\in \Gamma_\Phi}
%   }{\text{ext}}
%   \left[
%   \frac{1}{2}\;
%   \tn{\textsf{L}}_{h,\Delta t/2}^0
%   \left(
%   \bm\varphi^0_h,
%   \Phi^0_h,
%   \bm\varphi^{1/2}_h,
%   \Phi^{1/2}_h
%   \right)
%   + \frac{1}{2}\;
%   \tn{\textsf{L}}_{h,\Delta t/2}^1
%   \left(
%   \bm\varphi^{1/2}_h,
%   \Phi^{1/2}_h,
%   \bm\varphi^{1}_h,
%   \Phi^{1}_h
%   \right)
%   \right],\nonumber
% \end{align} 
% %
% where an explicit time dependence was included to account for the
% constraints imposed by the boundary conditions.
\begin{align}
\label{eq:11}
  \tn{\textsf{L}}_{h,\Delta
    t}^{01}&(\bm\varphi_h^0,\Phi_h^0,\bm\varphi_h^1,\Phi_h^1)
  =\\
&  \underset{\substack{
      \bm\varphi_a^{1/2} = \overline{\bm \varphi}(\bm
      X_a)\text{ if } \bm X_a\in \Gamma_\varphi\\
      \Phi_a^{1/2} = \overline{\Phi}(\bm
      X_a)\text{ if } \bm X_a\in \Gamma_\Phi}
  }{\text{ext}}
  \left[
%  \frac{1}{2}\;
  \tn{\textsf{L}}_{h,\Delta t/2}^0
  \left(
  \bm\varphi^0_h,
  \Phi^0_h,
  \bm\varphi^{1/2}_h,
  \Phi^{1/2}_h
  \right)
  +
% \frac{1}{2}\;
  \tn{\textsf{L}}_{h,\Delta t/2}^1
  \left(
  \bm\varphi^{1/2}_h,
  \Phi^{1/2}_h,
  \bm\varphi^{1}_h,
  \Phi^{1}_h
  \right)
  \right],\nonumber
\end{align} 
The resulting scheme is given in \S \ref{sec:expl-second-order}. A few remarks
are appropriate:
\begin{enumerate}
\item[(i)] Since the Hamiltonian of the system is non-separable, the
  obtained composition method does not coincide with Newmark's
  algorithm, as we noted earlier \cite{HairerEulerSymplectic,MataLew1}.
\item[(ii)] Another alternative to formulate an explicit, second-order variational time integrator is given by
$$\hat{\tn{F}}_{h,\Delta t}^{01}=\tn{F}_{h,\Delta t/2}^0\circ\tn{F}_{h,\Delta t/2}^{1},$$ 
namely, by inverting the order of the composition in (\ref{CompositionGeneral}) or, equivalently, inverting the order of the discrete Lagrangians $\tn{\textsf L}_{h,\Delta t}^{1}$ and $\tn{\textsf L}_{h,\Delta t}^{0}$ in (\ref{eq:11}). 
\item[(iii)] Numerical algorithms in position-momentum form require
  the specification of initial conditions in terms of the nodal values
  $(\Ne p_a(0),\tau_a(0))$. These can be computed with the help of
  (\ref{BCPositionMomentumForm}) from the
  initial configuration and the initial velocities and entropies $(\Ne
  v^0_a,\theta^0_a)$. We do this in the algorithm in \S \ref{sec:expl-second-order}.
\item[(iv)] A more compact algorithm could be possible by grouping some of
  the steps in the algorithm in \S \ref{sec:expl-second-order}, and a
  computer implementation could benefit from that. For the sake of
  clarity, in displaying the algorithm we kept precisely the steps
  involved from regarding the algorithm as the composition of the two
  first-order algorithms. We did obviate the computation of quantities
  in the midpoint of a step that were not needed later.
\end{enumerate}

\subsection{Second order with $\Delta t$ --- Non-autonomous case}
\label{sec:nonautonomous}

The construction of second-order integrators in time when the
functions $\overline{\bm\varphi}(\Ne X_a,t)$ and $\overline \Phi(\Ne
X_a,t)$ in \eqref{Dirichlet_1h} are other than constant needs some
further care \footnote{To be precise, they could be affine in time,
  and the algorithm would still be second-order.}. The essential
reason behind it is that higher order methods require a higher-order
sampling of these functions. It is not hard to see that in this case algorithms ${\tn{F}}_{h,\Delta t/2}^{0}$ and
${\tn{F}}_{h,\Delta t/2}^{1}$ cease to be adjoint to each other.

% For affine boundary conditions, the approximation of the
% velocities of the nodes lying along the Dirichlet boundary is
% exact. This is not the case in the non-affine case, and the resulting
% error is of first order, lowering the accuracy of the entire
% algorithm. It is simple to see how in this case algorithms
% ${\tn{F}}_{h,\Delta t/2}^{0}$ and ${\tn{F}}_{h,\Delta t/2}^{1}$ cease
% to be adjoint to each other.

The effect of the time-dependent Dirichlet boundary conditions is to
transform the mechanical system into one with a non-autonomous
Lagrangian. To formulate higher-order variational integrators for this
case, it is convenient to recover an autonomous Lagrangian by
including time as one more generalized coordinate of the system, see
\cite[Part 4]{Marsden2}. In this extended configuration manifold, an
integration algorithm maps $(t^k, \bm \varphi_h^k, \Phi_h^k,\Ne p^k,
\bm \tau^k)$ to $(t^k+\Delta t, \bm \varphi_h^{k+1}, \Phi_h^{k+1},\Ne
p^{k+1}, \bm \tau^{k+1})$. The key consequence of this perspective is
that time is regarded as a variable that is equivalent to $\bm
\varphi_h$ or $\Phi_h$, so in the symplectic-Euler-A method 
Dirichlet boundary conditions are evaluated at the beginning of the
time step, while this is done at the end of the time step in the
symplectic-Euler-B method.  Under these conditions, these two
algorithms are again adjoint to each other, so their composition
renders a second-order method, which is what is shown in the algorithm
\S \ref{sec:expl-second-order}. For constant (or in this case also affine) Dirichlet
boundary conditions in time, this algorithm coincides with that of \S
\ref{SecondOrderDeltaT}, but they differ otherwise. The difference
lies in that in this algorithm  the values {\it and} the
time derivatives of $\overline{\bm \varphi}$ and $\overline{\bm \Phi}$
are evaluated, while in the algorithm in \S \ref{SecondOrderDeltaT}
the time derivatives of  $\overline{\bm \varphi}$ and $\overline{\bm
  \Phi}$ are approximated by a simple finite difference formula. 

We shall not further expand the discussion on this aspect of the
algorithms here, since it requires a discussion of non-autonomous
systems and the introduction of further notation, and it is only
marginally related to the main goal of the paper.

\subsection{Explicit second-order algorithm $\hat {\tn F}_{h,\Delta
    t}^{10}$}
\label{sec:expl-second-order}
\noindent The precise description of the algorithm follows below.

\noindent \rule[-12pt]{\linewidth}{1pt}
\begin{algorithmic}[1]
\STATE Input: $(\bm\varphi_h^0, \Ne v_h^0,\Phi_h^0, \theta_h^0)$
\STATE Output: $\left(\bm \varphi_h^i,\Ne v_h^i,  \Phi_h^i,
  \theta_h^i\right)$ for $i=1,\ldots, M$
\STATE
\STATE \COMMENT{Initialization}
\STATE $\Ne p_a^0\gets m_{aa} \Ne v_a^0$ for all nodes
$a$
\STATE $\tau_a^0\gets \Upsilon_a(\bm \varphi_h^0, \theta_a^0)$ for all nodes $a$.
\STATE $\Delta t_{\text{eff}}\gets \Delta t/2$.
\STATE
\STATE \COMMENT{Advance in time}
\FOR{$k=0$ to $M-1$}
\STATE \COMMENT{First half, $F^0_{h,\Delta t_\text{eff}}$}
\FORALL {nodes $a$}
\IF { $\Ne X_a\in \Gamma_\Phi$}
%   \STATE $\Phi_a^{k+1/2} \gets \overline{\Phi}(\Ne X_a, t^k+\Delta  t_{\text{eff}})$
   \STATE $\theta_a^\text{pre} \gets \dot{\overline \Phi} (\Ne X_a, t^k)$ 
\ELSE
\STATE $\Phi_a^{k+1/2}\gets$ Solve for $\Phi_a^{k+1/2}$ from $\tau_a^k =
\Upsilon_a\left(\bm\varphi_h^k, (\Phi_a^{k+1/2}-\Phi_a^{k})/\Delta
t_{\text{eff}}\right) - 
\Delta t_{\text{eff}}\left[\tn H_a(\bm \varphi_h^k, \Phi_h^k) + \tn Q_a\right]$
\STATE $\theta_a^\text{pre} \gets (\Phi_a^{k+1/2}-\Phi_a^{k})/\Delta
t_{\text{eff}}$ 
\STATE $\tau_a^{k+1/2} \gets \Upsilon_a\left(\bm\varphi_h^k, \theta_a^\text{pre}\right)$
\ENDIF
\IF { $\Ne X_a\not\in \Gamma_\varphi$}
%\STATE $\bm\varphi_a^{k+1/2} \gets \overline{\bm\varphi}(\Ne X_a, t^k+\Delta t_{\text{eff}})$
%\ELSE
\STATE $\bm\varphi_a^{k+1/2} \gets \bm\varphi_a^k + \Delta t_{\text{eff}} \Ne p_a^k/m_{aa} -
\Delta t_{\text{eff}}^2\left[
\Ne S_a(\bm\varphi_h^k, \Phi_h^k, \theta_a^\text{pre}) -
  \Ne B_a(\bm\varphi_h^k)\right]/m_{aa}$.
\STATE $\Ne p_a^{k+1/2}\gets m_{aa} (\bm\varphi_a^{k+1/2}-\bm
\varphi_a^k)/\Delta t_{\text{eff}}$
\ENDIF

\ENDFOR
\STATE
\STATE \COMMENT{Second half, $F^1_{h,\Delta t_\text{eff}}$}
\FORALL {nodes $a$}
\IF { $\Ne X_a\in \Gamma_\varphi$}
\STATE $\bm\varphi_a^{k+1} \gets \overline{\bm\varphi}(\Ne X_a, t^{k+1})$
\ELSE
\STATE $\bm\varphi_a^{k+1} \gets \bm\varphi_a^{k+1/2} + \Delta t_{\text{eff}} \Ne p_a^{k+1/2}/m_{aa}$
\ENDIF
\ENDFOR
\FORALL {nodes $a$}
\IF { $\Ne X_a\in \Gamma_\Phi$}
   \STATE $\Phi_a^{k+1} \gets \overline{\Phi}(\Ne X_a, t^{k+1})$
   \STATE $\theta_a^\text{pre} \gets \dot{\overline \Phi} (\Ne X_a,
   t^{k+1})$ 
\ELSE
\STATE $\Phi_a^{k+1}\gets$ Solve for $\Phi_a^{k+1}$ from $\tau_a^{k+1/2} =
\Upsilon_a\left(\bm\varphi_h^{k+1}, (\Phi_a^{k+1}-\Phi_a^{k+1/2})/\Delta t_{\text{eff}}\right)$.
\STATE $\theta_a^\text{pre}\gets (\Phi_a^{k+1}-\Phi_a^{k+1/2})/\Delta t_{\text{eff}}$
\ENDIF
\ENDFOR
\FORALL {nodes $a$}
\IF { $\Ne X_a\in \Gamma_\varphi$}
\STATE $\Ne v_a^{k+1} \gets \dot{\overline{\bm\varphi}}(\Ne X_a, t^{k+1})$
\STATE $\Ne p_a^{k+1} \gets m_{aa} \Ne v_a^{k+1}$
\ELSE
\STATE $\Ne p_a^{k+1}\gets \Ne p_a^{k+1/2} 
%m_{aa} (\bm\varphi_a^{k+1}-\bm\varphi_a^{k+1/2})/\Delta
%t_{\text{eff}}
 - \Delta t_{\text{eff}} \left[\Ne S_a(\bm\varphi_h^{k+1}, \Phi_h^{k+1}, \theta_a^\text{pre}) -
  \Ne B_a(\bm\varphi_h^{k+1})\right]$
\STATE $\Ne v_a^{k+1} \gets \Ne p_a^{k+1}/m_{aa}$
\ENDIF
\IF { $\Ne X_a\in \Gamma_\Phi$}
\STATE $\theta_a^{k+1}\gets \theta_a^\text{pre}$
\STATE $\tau_a^{k+1}\gets \Upsilon_a\left(\bm\varphi_h^{k+1},
 \theta_a^{k+1}\right)$
\ELSE 
\STATE $\tau_a^{k+1} \gets \Upsilon_a\left(\bm\varphi_h^{k+1},
  \theta_a^\text{pre}\right)+ \Delta
t_\text{eff} \left[\tn H_a(\bm \varphi_h^{k+1}, \Phi_h^{k+1}) + \tn Q_a\right] $
\STATE $\theta_a^{k+1} \gets$ Solve for $\theta_a^{k+1}$ from
$\tau_a^{k+1} = \Upsilon_a\left(\bm\varphi_h^{k+1},
 \theta_a^{k+1}\right)$
\ENDIF
\ENDFOR
\ENDFOR
\end{algorithmic}
\noindent \rule{\linewidth}{1pt}
%\end{algorithm}

% \begin{algorithm2e}
% \caption{Composed method $\hat{\tn F}^{10}_{h,\Delta t}.$} \label{F10}
% \SetAlgoLined
% %\KwData{FE mesh $\ca T_h$, initial BC: $\{\bm\varphi_a^0,\theta_a\}_{a=1}^{n_t}$.}

% %Compute initial BC $\{\Ne p_a^0,\tau_a^0\}_{a=1}^{n_t}$ with (\ref{InitialBC_2}) and the mass matrix $\Ne m$\; 

% \ForAll{$\tn{time steps}$}{ 

% \ForAll{$\tn{ the nodes of the mesh}$}{
 
% First half-step\; 

% $\bm{\varphi}_a^{\frac 1 2}= 
% \bm{\varphi}_a^0 + 
% \dfrac{\Delta t}{2\tn m_a}
% \Ne p_a^0$\;  

% Solve $\tau_a^0=\Upsilon_a^{\left(\frac 1 2\right)-}
% \quad$ for $\quad\Phi_a^{\frac 1 2}$\; 

% $\Ne p^{\frac 1 2}_a=
% \dfrac{2\tn m_a}{\Delta t}
% \left(
% \bm\varphi_a^{\frac 1 2}-
% \bm\varphi_a^0
% \right)
% -\dfrac{\Delta t}{2}
% \Ne S_a^{\left(\frac 1 2\right)-}
% $\; 

% $\tau_a^{\frac 1 2}=
% \Upsilon^{\left(\frac 1 2\right)-}_a
% +\dfrac{\Delta t}{2}\tn H_a^{\frac 1 2}$\;

% Second half-step\; 

% Solve $\tau_a^{\frac 1 2}=\Upsilon_a^{\left(\frac 1 2\right)+}-
% \Delta t~\tn H_a^{\frac 1 2}\quad$ for $\quad\Phi_a^1$\;  

% $\bm{\varphi}_a^1= 
% \bm{\varphi}_a^{\frac 1 2} + 
% \dfrac{\Delta t}{2\tn m_a}
% \left(
% \Ne p_a^{\frac 1 2}
% -\dfrac{\Delta t}{2}\Ne S_a^{\left(\frac 1 2\right)+}
% \right)$\; 

% $\Ne p^1_a=
% \dfrac{2\tn m_a}{\Delta t}
% \left(
% \bm\varphi_a^1-
% \bm\varphi_a^{\frac 1 2}
% \right)
% $\; 

% $\tau_a^1=
% \Upsilon_a^{
% \left(\frac 1 2\right)+}$\;
%     }
%   }
% \end{algorithm2e}
%
%------------------------------------------------------------------------------------------------
%
\subsection{A note on the use of a nodal quadrature}

Explicit time integrators in solid dynamics require the use of a
lumped mass matrix. These are generally obtained through the GL quadrature
rule for the case of $P_1$ finite elements (see, e.g.,
\cite{Hughes1}). The  introduction of such quadrature does not
deteriorate the spatial approximation properties of the scheme for
linear elements
\cite{Cohen001,Durufle01}.

\subsection{Conservation properties of the algorithms}
\vskip .25cm
As in \S \ref{ConservedQuantitiesContinuum}, in this
section we assume that $\Gamma_{\bm \varphi}=\Gamma_\Phi=\emptyset$,  for simplicity.

\paragraph{Momentum conservation} One of the most attractive
properties of VIs is that the discrete flows exactly conserve the
quantities associated to symmetries of the discrete Langrangians, due
to a discrete analogue of Noether's theorem for continuous systems
\cite{Lew1,Lew2,ThesisAdrian,Marsden2}.  We briefly review here the
quantities conserved by this integrator, mostly focusing on the
thermal parts, which have not been widely discussed.

The standard conserved mechanical quantities follow from the fact that
the discrete Lagrangians proposed in (\ref{eq:6}) are invariant under
the action of rigid body translations and rigid body rotations when,
for example,  $\Ne
B=0, \tn Q=0, \Ne T = 0,$ and $\overline{ \textsf h}=0$, as the exact Lagrangian
does, see \S \ref{ConservedQuantitiesContinuum}. Therefore, the total
linear momentum and the total angular momentum are automatically
conserved by the algorithm. For completeness, the conserved quantities
are computed as
\begin{equation}
  \label{eq:29}
  \Ne L_h^k = \sum_{a=1}^N \Ne p_a^k \qquad \text{and} \qquad \Ne
  A_h^k= \sum_{a=1}^N \bm \varphi_a^{k} \times \Ne p_a^k,
\end{equation}
for the linear and angular momenta, respectively.

If $\textsf Q=0$ and $\overline {\textsf h}=0$, the algorithm also conserves the
total thermal momentum, namely, the value of the total entropy in the
system as computed from the discretization:
\begin{equation}
  \label{eq:20}
  \Xi_h^k = \sum_{a=1}^N \tau_a = \rho_0 \sum_{a=1}^N
  \sum_{\text{ring}(a)} w_{a,K}
  \eta(\nabla \bm \varphi_h^{k+1}|_K, \theta_a^{k+1})
\end{equation}
This follows from the symmetry of the discrete
Lagrangian with respect to rigid translations of the thermal
displacements. We briefly derive it next, essentially reproducing the
steps of the discrete Noether's theorem shown in, for example,
\cite{Lew1,Marsden2}. To this end, consider a one parameter group of
thermal displacements 
\begin{equation}
  \label{eq:21}
  \Phi^{k,\epsilon}_h = \Phi_h^k + \epsilon,
\end{equation}
where $\epsilon\in \mathbb R$ is a constant. This corresponds to
rigidly translating all thermal displacements by $\epsilon$. 
A simple computation shows that the discrete Lagrangians (\ref{eq:6}) are invariant under the actions of this set of translations, i.e., 
$$\tn{\textsf{L}}^{i}_{h,\Delta t}(\bm\varphi_h^k,\Phi_h^{k,\epsilon},
\bm\varphi_h^{k+1},\Phi_h^{k+1,\epsilon})=
\tn{\textsf{L}}^{i}_{h,\Delta t}(\bm\varphi_h^k,\Phi_h^k,
\bm\varphi_h^{k+1},\Phi_h^{k+1}),$$ 
for $i=0,1$, all $\epsilon\in \mathbb R$, and all $k=0,\ldots,M-1$, provided of course that  $\textsf Q=0$
and $\overline {\textsf h}=0$. 
This property in turn implies the invariance of $\tn{\textsf{L}}^{01}_{h,\Delta t}$ and the invariance of its corresponding discrete action sum $\tn{\textsf S}_d$. Therefore,
\begin{equation}
  \label{eq:22}
  \textsf L^{01}_{h,\Delta t}
  (\bm\varphi_h^k,\Phi_h^{k,\epsilon},\bm\varphi_h^{k+1},\Phi_h^{k+1,\epsilon})
  =   \textsf L^{01}_{h,\Delta t}
  (\bm\varphi_h^k,\Phi_h^{k},\bm\varphi_h^{k+1},\Phi_h^{k+1}).
\end{equation}
It then follows that
\begin{equation}
  \label{eq:23}
0 =   \frac{\partial \textsf S_d}{\partial \epsilon} \Big|_{\epsilon=0} =
\sum_{a=1}^N  D_{2a} \textsf L_{h,\Delta t}^{01}  (\bm \varphi_h^0,
\Phi_h^0, \bm \varphi_h^1, \Phi_h^1) +  \sum_{a=1}^N  D_{4a} \textsf L_{h,\Delta t}^{01}  (\bm \varphi_h^{M-1},
\Phi_h^{M-1}, \bm \varphi_h^M, \Phi_h^M),
\end{equation}
over a discrete trajectory $(\bm \varphi_h^0,\Phi_h^0,\ldots, \bm \varphi_h^M, \Phi_h^M)$, since by its definition  $\partial \textsf S_d/\partial \Phi_h^k=0$ for
$k=1,\ldots, M-1$. Using \eqref{Eq2} and \eqref{Eq4}, we can rewrite
this last equation as
\begin{equation}
  \label{eq:24}
  \sum_{a=1}^N \tau_a^0 = \sum_{a=1}^N \tau_a^M.
\end{equation}
The result in \eqref{eq:20} then follows from \eqref{eq:18}, and from
identifying $M$ with any intermediate time step $k$. 

\vskip .25cm
%
%----------------- Symplecticity
%
\paragraph{Symplecticity of the discrete flow}
The exact flow of a Lagrangian system conserves a symplectic bilinear
form, a straightforward consequence of its variational structure
\cite{Marsden2,Marsden3,Marsden4}. When discretized with a variational
integrator, the same symplectic form is conserved by the discrete flow
(see, e.g., \cite{Marsden2,Lew1}).  We do not reproduce the exact
conserved form here, since it is a straightforward application of
general results on variational integrators.  In the context of
continuum mechanics, a more general notion of symplecticity is also
important, the so-called multi-symplecticity. For details, we refer
the readers to \cite{Marsden3,Lew2}.

\comment{
To deduce an explicit expression for the symplectic form, first we define $\bm\Psi^k=(\varphi_1^k,...,\varphi_N^k,\Phi_1^k,...,\Phi_N^k)\in\RU^{4N}$ and consider a two-parameter set of initial conditions $(\bm\Psi^{\xi,\nu,0},\bm\Psi^{\xi,\nu,1})$ such that $\{\bm\Psi^{\xi,\nu,k}\}_{k=0}^{M}$ is resulting
discrete trajectory of the system. The corresponding variations are
$$\delta\bm\Psi^{\xi,k}=
\left.\dfrac{\partial}{\partial\nu}\bm\Psi^{\xi,\nu,k}\right|_{\nu=0},
\qquad
\delta\bar{\bm\Psi}^{\nu,k}=
\left.\dfrac{\partial}{\partial\xi}\bm\Psi^{\xi,\nu,k}\right|_{\xi=0},
\qquad
\delta^2\bm\Psi^k=
\left.\dfrac{\partial}{\partial\xi}\dfrac{\partial}{\partial\nu}
\bm\Psi^{\xi,\nu,k}\right|_{\xi,\nu=0},$$
and write $\delta\bm\Psi^k=\delta\bm\Psi^{0,k}$, $\delta\bar{\bm\Psi}^k=\delta\bar{\bm\Psi}^{0,k}$. 

Then, taking into account the symmetry of the mixed partial derivatives, we have that
$$\left.\dfrac{\partial}{\partial\xi}\right|_{\xi=0}
  \left.\dfrac{\partial}{\partial\nu}\right|_{\nu=0}
  \left(
  \tn{\textsf S}_d	
  \{\bm\Psi^{\xi,\nu,k}\}_{k=0}^{M-1}
  \right)-
  \left.\dfrac{\partial}{\partial\nu}\right|_{\nu=0}
  \left.\dfrac{\partial}{\partial\xi}\right|_{\xi=0}
  \tn{\textsf S}_d
  \left(	
  \{\bm\Psi^{\xi,\nu,k}\}_{k=0}^{M-1}
  \right)=0.$$
Repeating this computation for $\tn{\textsf L}_d(\bm\Psi^{\xi,\nu,0},\bm\Psi^{\xi,\nu,1})$ yields to another identity which may be replaced into the above equation to give
\begin{equation}
  D_{1j}D_{2i}\tn{\textsf L}_d(\bm\Psi^{M-1},\bm\Psi^M)
  (\delta\Psi^{M}_i\delta\bar\Psi^{M-1}_j-\delta\bar\Psi^M_i\delta\Psi^{M-1}_j)
  =
  D_{1j}D_{2i}\tn{\textsf L}_d(\bm\Psi^0,\bm\Psi^1)
  (\delta\Psi^1_i\delta\bar\Psi^0_j-\delta\bar\Psi^1_i\delta\Psi^0_j).
\end{equation}
The above antisymmetric bilinear form is the discrete Lagrangian symplectic form (evaluated $\delta\bm\Psi^k$ and $\delta\bm\bar\Psi^k$) which is preserved by the time evolution of the discrete system.
}
\paragraph{Energy conservation}
For an autonomous Lagrangian the total energy is exactly conserved
along its trajectories. Unfortunatelly, symplectic numerical methods
with constant time steps cannot be simultaneously symplectic and
energy conserving \cite{Ge001,Kane001}. Despite the above limitation,
symplectic integrators exhibit an excelent long term energy behavior
when applied to Lagrangian systems. More specifically, Theorem 8.1
\cite[Ch IX.8, pp.367 ]{Hairer1} states that the energy of the
numerical trayectories remains $\ca O(\Delta t^p)$ ($p$ the order of
the method) close to the exact one for exponentially long periods of
time if a small enough time step is adopted. Therefore, the first law of
thermodynamics is nearly exactly conserved by the discrete
trajectories.
%
%------------------- Stability -------------------------------------
%
\subsection{Stability}
As an ad-hoc stability criterion for the explicit algorithms proposed
here we have adopted that time steps are computed as a fraction of the
Courant condition 
%
%$$\Delta t\leqslant \dfrac{1}{4}\dfrac{h}{\Omega^\ast},$$
\begin{equation}\label{MaximumDt}
  \Delta t\leqslant \dfrac{h}{c_{\max}},
\end{equation}
where $c_{\max}$ is the largest value of $\omega/K$ in
(\ref{OmegaLinear}), for a linear thermo-elastic material. We have verified numerically that this
effectively behaves as the stability limit for this case.

%

%
%-------------------------------------------------------------------------------------------
%
\FloatBarrier
\section{Numerical examples}\label{SectionNumericalExamples}
In this section we present numerical evidence of the properties of the
algorithm $\hat{\tn{F}}_{h,\Delta t}^{10}$ from \S
\ref{sec:expl-second-order}. The  first example is devoted to show
that the algorithm converges quadratically with both the mesh size and
the time step. The second example is dedicated to show the
conservation properties of the algorithm when simulating the dynamics
of a nonlinear material.

\subsection{Convergence Properties of the Algorithm}
\label{sec:numerical-examples}

We begin by numerically observing the convergence rates of the
algorithm. To this end, we constructed an exact solution that depends
only on one of the spatial variables, so that we can simulate it and
compare. % in one, two, and three-dimensional cases. 
%We use the one-dimensional harmonic solutions from \S
%\ref{Example1Continuum}. More specifically, for a linear, isotropic
Such solution was obtained in \S \ref{Example1Continuum} for a linear
and isotropic thermo-elastic material, by proposing  harmonic
solutions of the form $\bm \varphi(\Ne X,t)=\Ne X+u(X,t)\Ne e_X$, with $u(X,t)=\Re[{A_\varphi \exp[i(K
X+\omega t)]}]$, and $\Phi(\Ne X,t)=\Re[{A_\Phi\exp[i(K X+\omega
t)]}]$. Non-trivial solutions exist only if \eqref{OmegaLinear} holds. This
implies that we require
\begin{equation}
  \label{eq:25}
  A_\Phi = A_\varphi \frac{E/\rho_0 -
   ( \omega/K)^2 }{ \gamma (\omega/K)}.
\end{equation}
%The elastic moduli of an isotropic linear elastic material are written
%as 
% \begin{equation}
%   \label{eq:33}
%   \Ne C = \lambda \Ne I\otimes \Ne I + 2\mu \mathbb I,
% \end{equation}
% where $\Ne I$ is the second-order identity tensor, $\mathbb I$ is
% the symmetric part of the fourth-order identity tensor, and $\lambda$
% and $\mu$ are the Lam\'e constants. For this example,
% $E=\lambda+2\mu$, where $E$ is the effective stiffness from \S~\ref{Example1Continuum}. 

In one spatial dimensions over a finite interval $\Omega=(0,L)$,
this is the solution of the initial value problem that satisfies
\eqref{eq:30} and the initial and boundary conditions
\begin{equation}
  \label{eq:26}
  \begin{aligned}
    u(X,0) & = \cos (K X), &   \qquad &&  \Phi(X,0) & = \frac{E/\rho_0 -
      ( \omega/K)^2 }{ \gamma (\omega/K)} \cos(K X),\\
    \dot{u}(X,0) & = -\omega \sin (K X), &   \qquad &&  \dot \Phi(X,0) & = -\frac{E/\rho_0 -
      ( \omega/K)^2 }{ \gamma (\omega/K)} \omega\sin(K X),\\
    u(0,t)  & = \cos(\omega t), &   \qquad &&  \Phi(0,t) & = \frac{E/\rho_0 -
      ( \omega/K)^2 }{ \gamma (\omega/K)} \cos(\omega t),\\
    u(L,t)  & = \cos(K L + \omega t), &   \qquad &&  \Phi(L,t) & = \frac{E/\rho_0 -
      ( \omega/K)^2 }{ \gamma (\omega/K)} \cos(K L +\omega t),
  \end{aligned}
\end{equation}
for all $X\in (0,L)$ and all $t>0$.

Out of the four possible values for $\omega$ in \eqref{OmegaLinear},
we chose $\omega_{++}$, so that $\omega$ is real. We set the material
properties to $E=20$, $\gamma=0.1$, $c=0.1$, $\rho_0=1$, $\kappa=0.1$,  $\theta_0=10$, and $\eta_0=0$,
in standard SI units. Therefore,
$c_{\max}=\omega_{++}/K=4.67379$. For this example we set
$\omega_{++}=4$ and $A_\varphi=1$, and computed $K=0.855837$ and
$A_{\Phi}=-3.94603$. The length of the domain is $L=100$, and the
domain was meshed with equal affine elements (two-node segments) with mesh sizes
$h\in\{10,5,2.5,1.25,0.625,0.3125,0.15625\}$. The corresponding time steps were
selected in order to respect (\ref{MaximumDt}), and were set to $\Delta
t\in\{0.5,0.25,0.125,0.0625,0.03125,0.15625,0.078125\}$. 

For a given time $t^k$, the error $e_{h,\Delta t}(t^k)$ associated with a
particular mesh size and time step  is computed
as the relative $L_2$-norm of the difference between some field  $z_{h,\Delta t}^k(\Ne X)$ of the discrete 
solution and the corresponding field $
z^\ast(\Ne X,t^k)$ of the exact solution, namely
\begin{eqnarray}
  e_{h,\Delta t}(t^k)
  &=&
  \dfrac{1}{z^\ast(t^k)}
  \sum_{e\in \ca{T}_h}\left(\sum_{p=1}^{N_q}w_p
  \left|
   z_{h,\Delta t}^k(\bm\xi_p)-
   z^\ast(\bm\xi_p,t^k)
  \right|^2\; dV\right)^{\frac 1 2}
  \\
  &\approx &
  \dfrac{1}{z^\ast(t^k)}
  \left(\int_\Omega
  \left|
  z_{h,\Delta t}^k(\Ne X)- z^\ast(\Ne X,t^k)
  \right|^2\;dV\right)^{\frac 1 2},
  \nonumber
\end{eqnarray}  
where $z^\ast(t)=(\int_\Omega z^{\ast}(\Ne X,t)^2\;
dV)^{1/2}$, and $\{w_p,\bm\xi_p\}_{p=1}^{N_q}$ is the Gauss-Legendre
quadrature rule. %, and $\Vert\cdot\Vert$ is the Euclidean norm in
%$\RU^{d+1}$.  
Here $z^\ast$ is one of the mechanical or thermal positions or velocities.
%Specifically, for the positions $\bm z^\ast = (\bm \varphi,\Phi)$, while for the
%velocities $\bm z^\ast=(\dot{\bm \varphi}, \theta)$.

\begin{center}
\begin{figure}[ht]
  \begin{minipage}[ht]{\textwidth}
  \centering
  \includegraphics[width=0.8\textwidth]{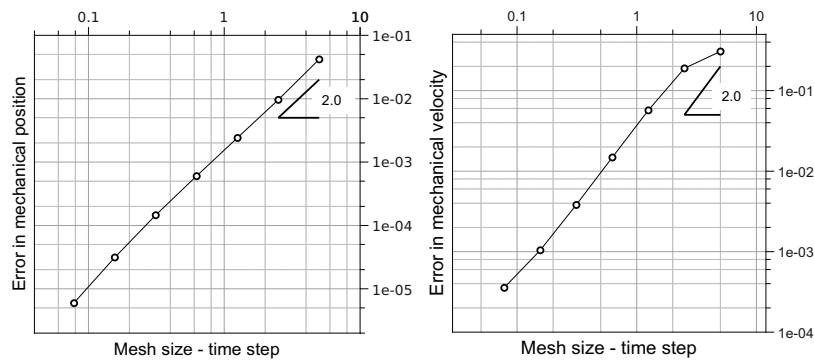}
  \end{minipage}
  \caption{Convergence rate of the algorithm measured in terms of the
    $L^2$-norm of the error in the mechanical (position and velocity) fields. }
  \label{ExampleAnalytic1}
\end{figure} 
\end{center}

\begin{figure}[ht]
  \begin{minipage}[ht]{\textwidth}
  \centering
  \includegraphics[width=0.8\textwidth]{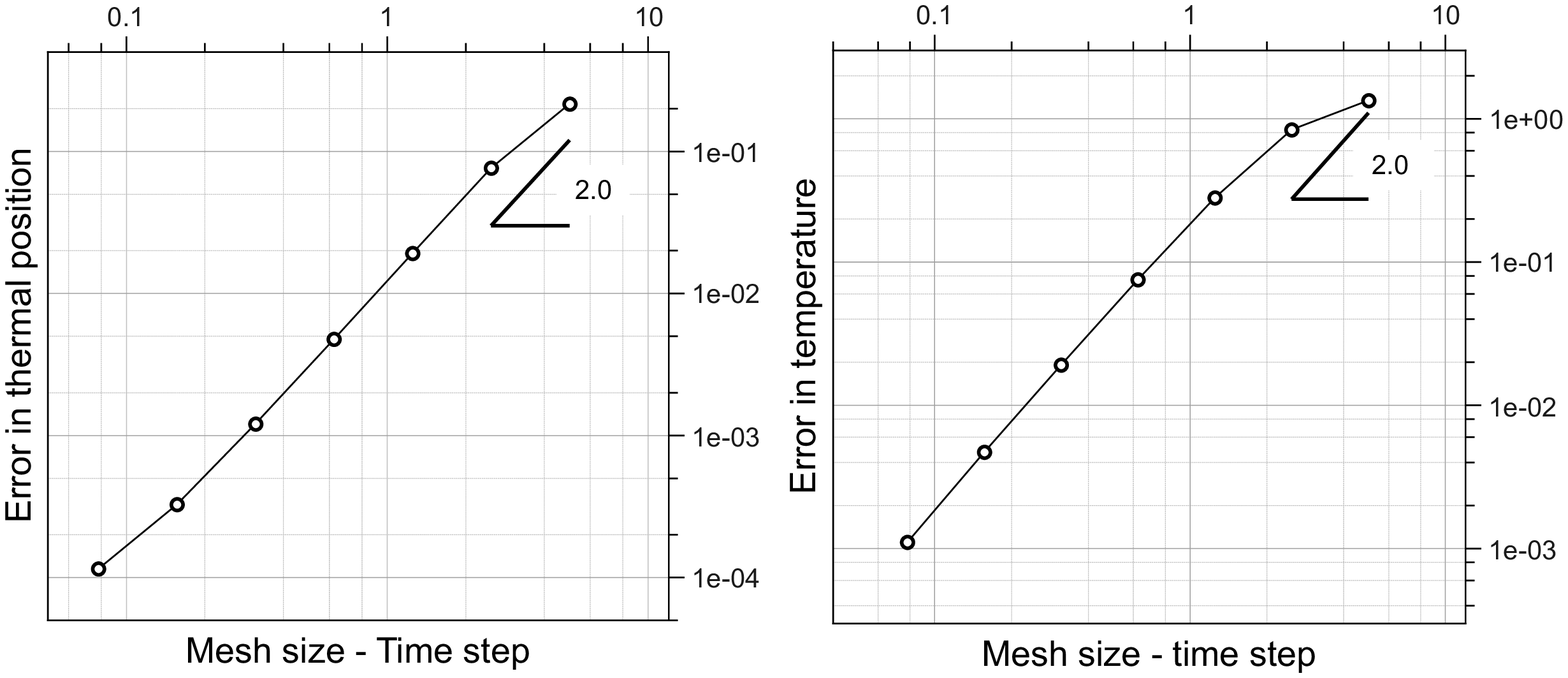}
  \end{minipage}
  \caption{Convergence rate of the algorithm measured in terms of the
    $L^2$-norm of the error in the thermal (position and velocity) fields. }
  \label{ExampleAnalytic2}
\end{figure}

\begin{figure}[ht]
  \begin{minipage}[ht]{\textwidth}
  \centering
  \includegraphics[width=0.9\textwidth]{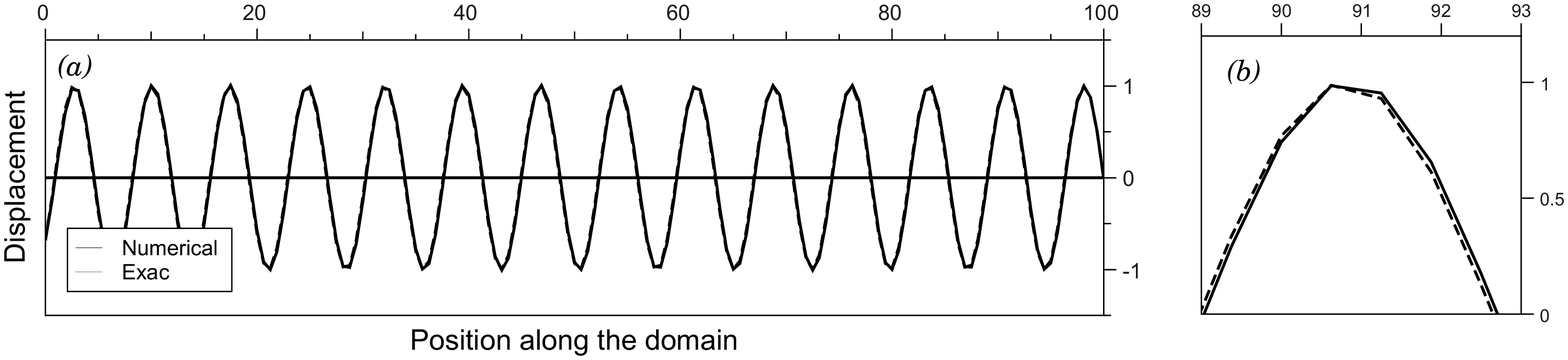}
  \end{minipage}
  \caption{$a)$ Exact and computed displacement field for one of the
    meshes we used. An enlarged view to observe the difference
    between the exact and the numerical solutions is also shown in
    $(b)$.  }
  \label{Zoom1}
\end{figure}

\begin{figure}[ht]
  \begin{minipage}[ht]{\textwidth}
  \centering
  \includegraphics[width=0.9\textwidth]{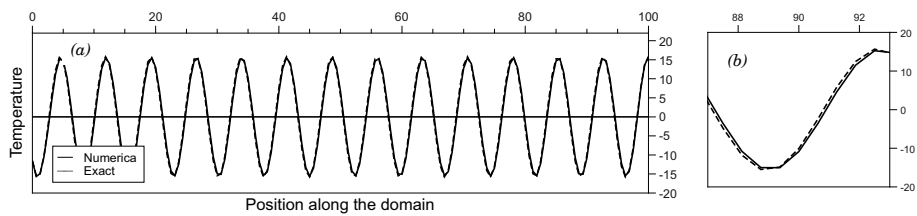}
  \end{minipage}
  \caption{$a)$ Exact and computed temperature field for one of the
    meshes we used. An enlarged view to observe the difference
    between the exact and the numerical solutions is also shown in $(b)$.  }
  \label{Zoom2}
\end{figure}

Figures \ref{ExampleAnalytic1} and \ref{ExampleAnalytic2} present the
evolution of $e_{h,\Delta t}(1)$ with the pairs $(h,\Delta t)$. The
algorithm exhibits a quadratic rate of convergence for both the error
computed for the positions ($\bm\varphi,\Phi$) and the error computed
for the velocities ($\dot{\bm \varphi},\theta$). %
Figures \ref{Zoom1} and \ref{Zoom2} present snapshots of the exact and
the numerical solutions for displacement and temperature at $t=1$. In
both cases, a zoomed out view of a smaller zone in the domain is also shown.


\subsection{Dynamics of a thermo-elastic three-dimensional body}
In this example we simulate the dynamics of three-dimensional
thermo-elastic beam that moves free of external mechanical or thermal
forces. In this way, the energy, linear and angular momenta, and
entropy of the beam have to conserved through the motion. Thus, this
example serves to illustrate the conservation properties of the
algorithm.

\begin{figure}[ht]
  \begin{minipage}[ht]{1\textwidth}
  \centering
  \includegraphics[width=.5\textwidth]{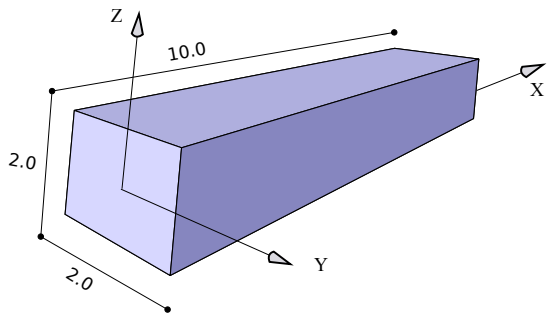}
  \end{minipage}
  \caption{Three-dimensional view of the reference configuration for
    the thermo-elastic beam in  this example.}
  \label{Example3D_1}
\end{figure}

\begin{figure}[ht]
  \begin{minipage}[ht]{1\textwidth}
  \centering
  \includegraphics[width=.95\textwidth]{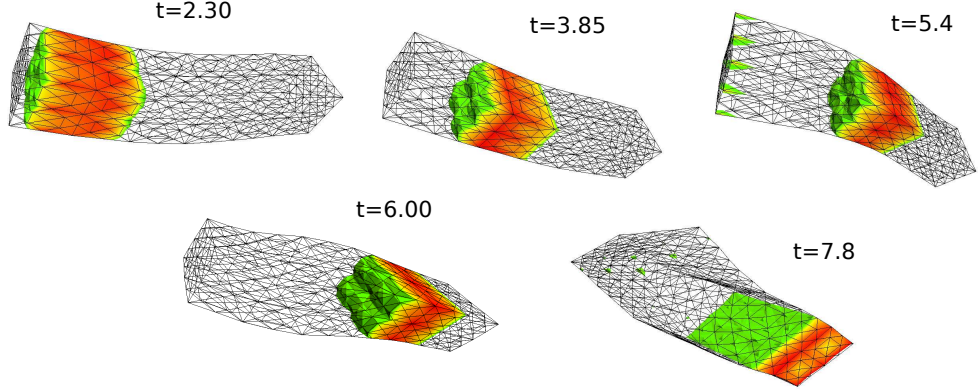}
  \end{minipage}
  \caption{Sequence of snapshots during the free-fly phase of this
    numerical example, showing the deformed configuration of the beam
    and contours highlighting the parts of the beam with temperatures
    in the range $[10.5,15]$.}
  \label{Example3D_2}
\end{figure}

The geometric description of the beam is depicted in
Fig. \ref{Example3D_1}. It was discretized with 4608 affine
tetrahedral elements, resulting in a total of 1061 nodes, see Fig. \ref{Example3D_2}.
The beam was assumed to be made of a material following the nonlinear
constitutive relation (\ref{NonLinearConstitutiveEquation}), with
$\rho_0=1.5$, $\mu=83.33$, $\lambda=55.55$, $\gamma=0.5$,
$\theta_0=10$, $c=5$, $\eta_0=10$ and $\kappa=1$, all of them in
standard SI units.

To initiate the motion of the beam, we load it both mechanically and
thermally for a short period of time, namely, 2.0 s. After that, the
beam can move being traction-free and entropy-flux free on its entire
boundary. Body forces and entropy sources per unit mass are absent as
well throughout the simulation. The mechanical and thermal load during
the first 2.0 s of the simulation is accomplished by setting
\begin{enumerate}
\item[(i)] $\Phi(\Ne X, t)=\theta_0t+\frac{40}{3}\sin(\frac{3}{10} t)$ if $\Ne
  X\equiv(X,Y,Z)\in \{0\}\times [-1,1]\times[-1,1]$, 
\item[(ii)] $\bm\varphi(\Ne X,t)=(X-\frac{1}{4}t,Y-\frac{3}{2}t,Z+\frac{4}{5}t)$ if $\Ne X\equiv (X,Y,Z)\in
  \{10\}\times [-1,1]\times[-1,1]$,
\item [(iii)] Zero entropy flux wherever $\Phi$ is not specified, and
\item [(iv)] Traction free wherever $\bm \varphi$ is not specified.
\end{enumerate}  
All constants above are, again, prescribed in appropriate SI
unites. The time integration of this problem was performed with
algorithm $\hat{\tn{F}}_{h,\Delta t}^{10}$ and a time step $\Delta
t=0.0025$ s.
    
The resulting dynamics of the beam is highly nonlinear involving
forced vibrations due to the action of biaxial bending, twisting,
shearing and compression that are simultaneously coupled with the
nonlinear propagation of thermal waves. Fig. \ref{Example3D_2} shows
a sequence of snapshots of the deformed shaped of the beam and
volumetric contours highlighting the parts
of the beam with temperatures in the range $[10.5,15]$. As expected,
the presence of thermal waves is clearly seen in the numerical solution.

\begin{figure}[ht]
  \begin{minipage}[ht]{1\textwidth}
  \centering
  \includegraphics[width=\textwidth]{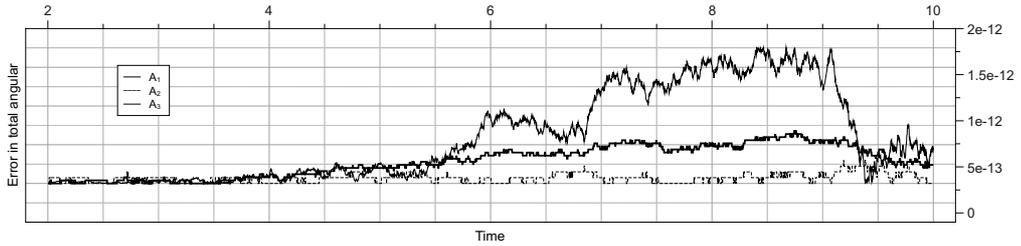}
  \end{minipage}
  \caption{Error in the components of the total angular momentum $\Ne A$.}
  \label{Example3D_3}
\end{figure}

\begin{figure}[ht]
  \begin{minipage}[ht]{1\textwidth}
  \centering
  \includegraphics[width=\textwidth]{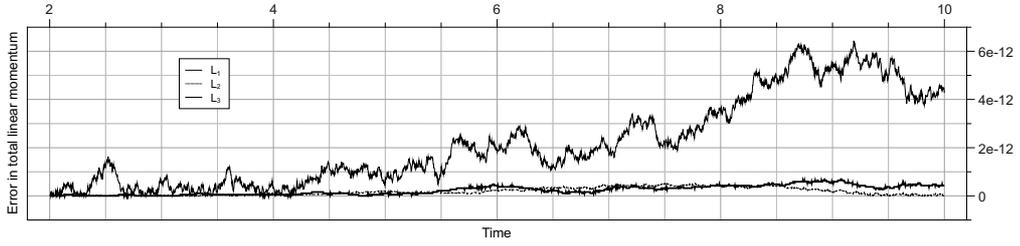}
  \end{minipage}
  \caption{Error in the components of the total linear momentum $\Ne L$.}
  \label{Example3D_4}
\end{figure}

\begin{figure}[ht]
  \begin{minipage}[ht]{1\textwidth}
  \centering
  \includegraphics[width=1\textwidth]{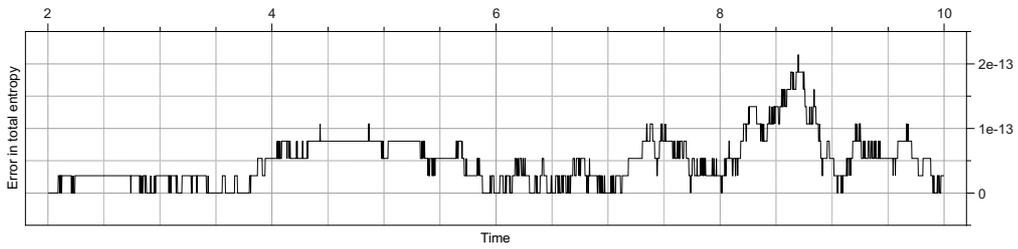}
  \end{minipage}
  \caption{Error in the total entropy.}
  \label{Example3D_5}
\end{figure}

\begin{figure}[ht]
  \begin{minipage}[ht]{1\textwidth}
  \centering
  \includegraphics[width=\textwidth]{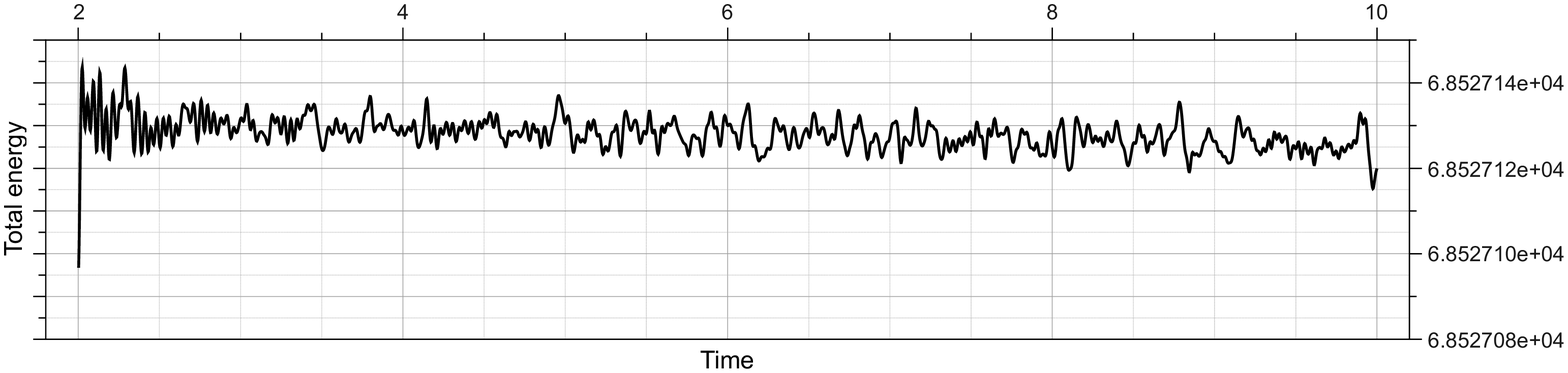}
  \end{minipage}
  \caption{Time evolution of the total energy.}
  \label{Example3D_6}
\end{figure}

The conservation properties of the proposed algorithm are verified in
Figs. \ref{Example3D_3}, \ref{Example3D_4} and \ref{Example3D_5},
where it is seen that the total linear ($\Ne L$, computed with \eqref{eq:29}) and angular momentum
($\Ne A$, computed with \eqref{eq:29}), and the total entropy ($\Xi$, computed with \eqref{eq:20}
at each time step),
are conserved up to round-off error after the external loading has
ceased ($t>$2.0 s). The specific values that are conserved, as
determined at $t=2.0$ s, are
\begin{eqnarray*}
  \Ne L &=& (31.85271482947, 
           -351.2418171388,
            187.4483240957 )\\
  \Ne A&=&(   0.760167559539, 
          -1337.826072699, 
          -2508.077227983 )\\
\Xi &=&3737.827302656
\end{eqnarray*}
where all the digits of the values that remain unchanged during the
simulation are shown. The error in every component of those quantities
in all the cases smaller than $10^{-11}$.

The energy of the system $\textsf H_h$, from \eqref{eq:28}, is not
exactly conserved. However, as it is typical of variational or
symplectic methods (see, e.g. \cite{Hairer1}), it is nearly exactly conserved, with small
oscillations around a value that is essentially conserved for
exponentially long times as the time step decreases. This is what we
observe in  Fig. \ref{Example3D_6}.

%
%--------------------------------------------------------------------------------------------
%
\section{Conclusions}\label{SectionConclusions}

The Hamiltonian structure of Green and Naghdi theory of type II for
thermo-elastic bodies with finite speed thermal waves enables the
construction of variational integrators over finite element
discretizations of the continuum problem.  The resulting methods are  said to be consistent with the laws of
thermodynamics in the sense that, for adiabatic systems, (i) they nearly
exactly conserve the total energy for an exponentially long period of
time (first law), and (ii) they exactly conserve the total entropy of
the system (second law). Moreover, they exactly conserve the total
linear momentum and the total angular momentum of the system.

The challenge of formulating variational integrators that are
thermodynamically consistent for  bodies undergoing
irreversible processes, such as heat conduction of Fourier type or
inelastic constitutive relations, remains. Thermodynamic consistency
in this case would amount to guaranteeing that for isolated systems
the energy will be (nearly) conserved and that the entropy of the
system will never decrease.

%
%----------------------------------------------------------------------------------------------------
%
\appendix
\section{Explicit expressions of derivatives of $\textsf L_h$}\label{sec:expl-expr-deriv}
The values of the derivatives of ${\sf L}_h$ from \S
\ref{sec:discretization-space} needed for the general algorithm in \S
\ref{sec:discr-time:-vari} are
\begin{equation*}
\begin{aligned}
  \Ne S_a^{i,k} &=
\frac{\partial \textsf W_{h}}{\partial \bm \varphi_a}
\left(\bm\varphi_h^{k+i},\Phi_h^{k+i},\frac{\Phi_h^{k+1}-\Phi_h^k}{\Delta
  t}\right) =\sum_{q=1}^{N_q} w_q \Ne P \left(\nabla \bvarphi_h^{k+i}(\bm
\xi_q), \nabla \Phi_h^{k+i}(\bm \xi_q),
\frac{\Phi_h^{k+1}-\Phi_h^k}{\Delta t}(\bm \xi_q)\right) \nabla \tn
N_a(\bm \xi_q), %\label{eq:Scomp}
\\
\Ne B_a^k &= -\frac{\partial \textsf E_h}{\partial
  \bm\varphi_a}(\bm\varphi_h^k,\Phi_h^k) =
-\sum_{q=1}^{N_q}w_i\rho_0\frac{\partial V_B}{\partial
  \bvarphi}(\bvarphi_h^{k}(\bm{\xi}_q)) \tn N_a(\bm \xi_q)
+
  \sum_{j=1}^{N_{t}}w_j^t \Ne T\; \tn N_a(\bm \xi^t_j), %\label{eq:Bcomp}
\\
\textsf
Q_a^k&=-\frac{\partial \textsf E_h}{\partial \Phi_a}(\bm
\varphi_h^k,\Phi_h^k) = \sum_{q=1}^{N_q}w_q\rho_0 \tn {\textsf{Q}}\;
  \tn N_a(\bm \xi_q)
+  \sum_{j=1}^{N_{h}}w_j^h
  (\overline{\tn h}\; \tn N_a)(\bm \xi_j^h), %\label{eq:Qcomp}
\\
  \Upsilon_a^{i,k}& = -\frac{\partial\textsf W_h}{\partial
    \theta_a}\left(\bm\varphi^{k+i}_h,\Phi_h^{k+i},\frac{\Phi_h^{k+1}-\Phi_h^k}{\Delta
    t}\right)=  \rho_0\sum_{q=1}^{N_q}w_q
  \eta\left(\nabla\bm\varphi_h^{k+i}(\bm\xi_q),
%  \nabla\Phi_h^{k+i}(\bm\xi_q),
\frac{\Phi_h^{k+1}-\Phi_h^k}{\Delta
    t}(\bm\xi_q)\right) \tn N_a(\bm \xi_q),%\label{eq:Upsiloncomp}
\\
\tn H_a^{i,k} &=
-\frac{\partial \textsf W_{h}}{\partial \Phi_a} \left(\bm
\varphi_h^{k+i},\Phi_h^{k+i},\frac{\Phi_h^{k+1}-\Phi_h^k}{\Delta
  t}\right) =  \sum_{q=1}^{N_q} w_q \Ne h\left(\nabla\bm\varphi_h^{k+i}(\bm\xi_q),
  \nabla\Phi_h^{k+i}(\bm\xi_q)
%,\frac{\Phi_h^{k+1}-\Phi_h^k}{\Delta    t}(\bm\xi_q)
\right) \nabla\tn N_a(\bm \xi_q),%\label{eq:Hcomp}
\end{aligned}
\end{equation*}
where we have used the assumption that $\eta$ does not depend on $\bm
\beta$ and that $\Ne h$ does not depend on $\theta$.

\end{document}